\def\EDGARclean{1}

\documentclass[acmtog,nonacm]{acmart} 

\AtBeginDocument{%
  }

\setcopyright{none}
\acmJournal{TOG}
\acmDOI{}
\acmISBN{}

\acmSubmissionID{}

\usepackage{graphicx}
\usepackage{array}
\usepackage{booktabs}
\usepackage{makecell}
\usepackage{multirow}
\usepackage[table]{xcolor}
\usepackage{hyperref}
\newif\ifshowrevision
\ifdefined\EDGARclean
  \showrevisionfalse
\else
  \showrevisiontrue
\fi
\definecolor{revisionblue}{RGB}{0,70,173}
\definecolor{revisionorange}{RGB}{230,110,0}
\ifshowrevision
  \newcommand{\newtext}[1]{\textcolor{black}{#1}}
  \newcommand{\revisedtext}[1]{\textcolor{black}{#1}}
  \newcommand{\newmaterialcolor}{\color{black}}
\else
  \newcommand{\newtext}[1]{\textcolor{black}{#1}}
  \newcommand{\revisedtext}[1]{\textcolor{black}{#1}}
  \newcommand{\newmaterialcolor}{\color{black}}
\fi

\begin{document}

\title{EDGAR : Efficient Diffusion-Driven Gaussian Avatar Refinement via Synthetic Failure}

\author{Kwan Yun}
\email{yunandy@kaist.ac.kr}
\orcid{0000-0002-0205-5203}
\affiliation{%
  \institution{KAIST, Visual Media Lab}
  \city{Daejeon}
  \country{South Korea}
}

\author{Minhyeok Bang}
\affiliation{%
  \institution{KAIST, Visual Media Lab}
  \city{Daejeon}
  \country{South Korea}
}

\author{Colette Suhjung Koo}
\affiliation{%
  \institution{KAIST, Visual Media Lab}
  \city{Daejeon}
  \country{South Korea}
}

\author{Junyong Noh}
\email{junyongnoh@kaist.ac.kr}
\orcid{0000-0003-1925-3326}
\affiliation{%
  \institution{KAIST, Visual Media Lab}
  \city{Daejeon}
  \country{South Korea}
}

\renewcommand{\shortauthors}{Yun et al.}

\ccsdesc[500]{Computing methodologies~Image processing}
\ccsdesc[300]{Computing methodologies~Animation}
\ccsdesc[300]{Computing methodologies~Computer vision}
\keywords{Gaussian Avatar, Image Refinement, Diffusion model}
\begin{teaserfigure}
    \centering
    \vspace{-2mm}
  \includegraphics[width=\textwidth]{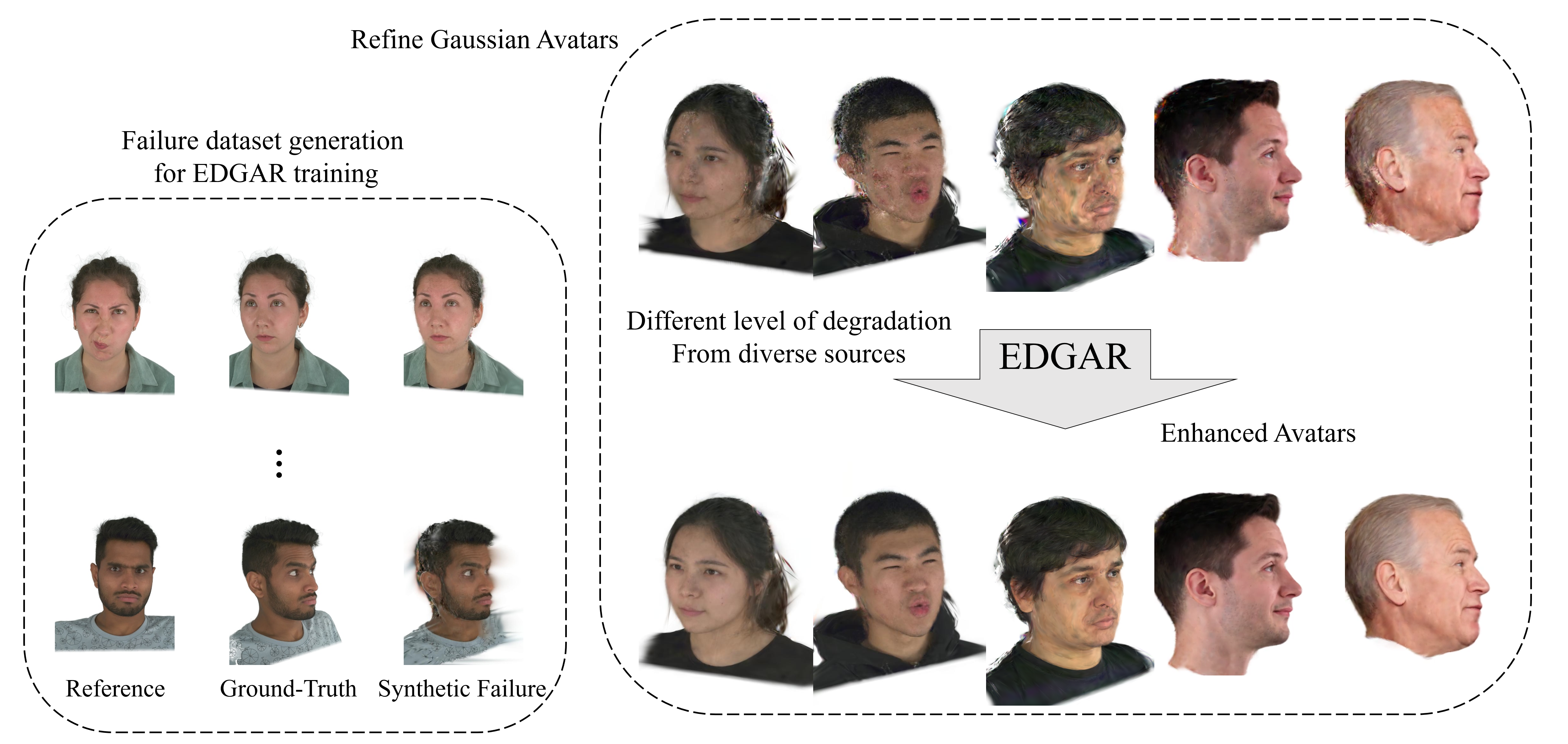}\vspace{-1mm}
  \caption{From diverse failure cases of Gaussian avatars, EDGAR learns to refine arbitrary avatars trained with a video or an image, acting as both an artifact remover and a quality enhancer.}
  \Description{}
  \label{fig:teaser}
\end{teaserfigure}


\newcommand{\fixq}[1]{\textcolor{black}{#1}}

\begin{abstract}
Recent advances in Gaussian splatting have enabled efficient and photorealistic head avatar reconstruction, yet monocular video–based or image-based methods still suffer from under-constrained geometry and texture, leading to artifacts such as blurring, stretching, and temporal instability. We present \textbf{EDGAR} (Efficient Diffusion-Driven Gaussian Avatar Refinement), a conditional diffusion framework designed to enhance the quality of Gaussian avatars by correcting artifacts while preserving identity fidelity. To train EDGAR, we construct EDGAR-SET, a hybrid triplet dataset composed of real-captured and synthetic-rendered samples, which simulates realistic Gaussian avatar failure cases by pairing degraded renderings with ground-truth and reference images. EDGAR refines degraded frames through a fusion block that injects identity features extracted by the self-supervised Sapiens model into a one-step diffusion backbone, enabling effective artifact removal without sacrificing efficiency. Our EDGAR model incorporates a Mixture-of-Experts (MoE) architecture to address diverse degradation types and severity levels. \newtext{The refined multi-view renderings are subsequently used as pseudo-targets to optimize a single underlying Gaussian avatar, so that the final avatar can be rendered without applying EDGAR to every output frame.} \revisedtext{Experiments on three different Gaussian avatar reconstruction methods, each evaluated under three different settings, show improvements in both perceptual and reconstruction measures over the raw outputs and an SDEdit-style refinement baseline.} \newtext{In addition, cross-view evaluation demonstrates improved calibrated consistency, while tests under large yaw and open-mouth expressions verify robustness under extreme settings.} \revisedtext{These results collectively indicate a useful operating regime for recoverable appearance artifacts.} The code is publicly available at \href{https://kwanyun.github.io/edgar/}{EDGAR}.
\end{abstract}


\maketitle

\section{Introduction}
\label{sec:intro}
Personalized head avatars are now integral to social gaming, VR telepresence, and mixed-reality interfaces, where users demand both photorealism and low latency.  Mesh-based reconstruction pipelines~\cite{feng2021learning,lattas2020avatarme} offer the advantage of compact representations and practical usability thanks to their fixed mesh structure. However, reliance on meshes makes it difficult to capture high-frequency details and to scale to high-resolution geometry, particularly in regions such as hair. In contrast, recent efforts built on volumetric neural-radiance fields (NeRF)~\cite{mildenhall2021nerf} have achieved impressive view synthesis results. Unfortunately, these methods require heavy per-scene optimization and remain slow for interactive use ~\cite{zielonka2023instant,zheng2022avatar,kirschstein2023nersemble}. Similarly, GAN-driven work achieved striking 3D-aware imagery~\cite{li2023efficient,an2023panohead,ye2023geneface}, yet encoded geometry only implicitly, limiting direct control over facial animation. These achievements have established the visual-quality bar while exposing the need for representations that are simultaneously lightweight, editable, and real-time ready.

The field has since pivoted to rasterizable 3D Gaussian splats ~\cite{kerbl20233d}, which replace dense ray-marching with a set of anisotropic blobs that are both GPU-friendly and easily deformed. Multi-view pipelines~\cite{xu2024gaussian,giebenhain2024npga,hu2024gaussianavatar} reach fast render times while faithfully tracking the face and producing realistically synthesized 
expressions. Monocular-video ~\cite{qian20243dgs,chen2024monogaussianavatar,zhang2025hravatar,chen2025mixedgaussianavatar} methods lower the capture barrier for further usability. Together, these Gaussian-based methods retain the photorealism pioneered by NeRF systems while unlocking interactive performance and casual acquisition, bringing personalized avatars within reach of everyday applications.

Nevertheless, compressing capture into a monocular video still leaves the reconstruction problem severely under-constrained. Monocular pipelines must infer geometry and texture in unobserved regions, which often causes artifacts—such as texture stretching, floating hair, or depth ambiguities—when rendering side profiles, wide-open mouths, or raised eyebrows. These issues are largely absent in avatars reconstructed from dense camera arrays, whereas limited observations in monocular settings leave unobserved regions under-constrained. While a few recent methods utilize diffusion-based image generation to fill this gap by adding and removing noise using pretrained generative models~\cite{meng2021sdedit} to refine the output~\cite{wu2025text,wu2024portrait3d}, these approaches inherently involve guessing appearance from a limited prior, thus failing to reliably
preserve identity details, especially in unseen poses and expressions.

To close this quality gap, we address avatar enhancement through a diffusion-prior-based refinement framework, which we term Efficient Diffusion-Driven Gaussian Avatar Refinement (EDGAR). Diffusion-based refinement is particularly well suited to this problem because strong generative priors can recover plausible high-frequency details while preserving overall visual realism under rendering degradation. The core idea is straightforward: given a rendered frame from a noisy or blurry Gaussian avatar and a reference image of the subject, EDGAR learns to refine the output into a sharper and more faithful rendering. This enables us to take an arbitrary Gaussian avatar—even when trained under suboptimal monocular conditions—and upgrade it to higher visual quality. However, learning this mapping is non-trivial: there is no existing dataset of "imperfect avatar vs. ground-truth render" pairs, and the appearance of a degraded avatar is not uniquely defined.

To address this, we construct EDGAR-SET, a hybrid dataset that combines real human data with synthetically generated samples to increase diversity in both identity and viewpoints. The dataset is organized as triplets: (1) rendered images from degraded avatars, (2) corresponding ground-truth images, and (3) reference identity images, as shown on the left of Figure~\ref{fig:teaser}. The synthetic component enables controlled simulation of realistic failure modes (e.g., sparse-view training, unseen poses, and expression variations), while the human dataset grounds the model in real-world appearance distributions. This combination ensures broad coverage over identity,
pose, and rendering artifacts.

EDGAR learns a conditional refinement function by injecting reference features~\cite{khirodkar2024sapiens} into a single-step diffusion backbone~\cite{sauer2024adversarial}. To robustly handle diverse failure modes, we further adopt a Mixture-of-Experts (MoE) architecture, allowing the model to specialize across different types and severities of degradation. In addition, \revisedtext{we use an expression consistency loss to encourage preservation of facial dynamics and subtle muscle movements during refinement.} \revisedtext{Our results demonstrate that EDGAR improves overall rendering fidelity and perceptual quality.}

\newtext{Our contributions are summarized as follows:}
\begin{itemize}
    \item \newtext{We construct EDGAR-SET, a hybrid collection of paired Gaussian-avatar failures, ground-truth targets, and identity references designed around recoverable head-avatar artifacts.}
    \item \newtext{We introduce EDGAR, an efficient diffusion-driven refinement framework that learns to correct rendered artifacts of animatable Gaussian head avatars and distills the refined 2D targets back into the underlying 3D avatar.} 
    \item \newtext{We develop subject-reference conditioning and expression-aware supervision to preserve identity and facial motion during refinement, together with a lightweight MoE module that handles diverse degradation patterns without increasing the final avatar rendering cost.}    
\end{itemize}

\section{Related Work}

\subsection{Head Reconstruction}
Classical head reconstruction leverages priors from 3D morphable models (3DMMs)~\cite{blanz1999morphable, li2017learning, cao2013facewarehouse} to recover identity and expression from images or videos. Unlike video based retargeting~\cite{guo2024liveportrait,zhao2025x}, 3D reconstruction provides explicit geometric and appearance representations, enabling more controllable and physically grounded avatar generation. Building on 3DMM priors, many methods tackle monocular reconstruction and tracking with varying levels of geometric detail and expression, either through classical fitting~\cite{blanz2003face,romdhani2005estimating,cao2014displaced,thies2016face2face} or, more recently, learning-based methods~\cite{zhu2016face,danvevcek2022emoca, deng2019accurate}. While these methods are efficient and provide explicit control through constructed blendshape parameters, their expressiveness in both dynamic geometry and subtle details is limited due to the constraints of the constructed space of defined 3DMM.

To enhance quality beyond the linear expressiveness of 3DMMs, recent approaches adopt implicit neural representations such as SDF~\cite{park2019deepsdf}, OccupancyNet~\cite{mescheder2019occupancy}, NeRF \cite{mildenhall2021nerf}, and Instant-NGP~\cite{muller2022instant}. Thanks to their flexibility, these representations enable effective reconstruction of fine geometric details such as hair wisps and beards~\cite{giebenhain2023learning, zheng2022imface, zheng2022avatar,zielonka2023instant,zheng2023pointavatar}. These advances have substantially improved the realism of head avatars, but often at the cost of high computational complexity or slow rendering speed, which limits their practicality in interactive applications.

\subsection{Gaussian Head Avatars}
Recently, 3D Gaussian Splatting (3DGS)~\cite{kerbl20233d} has emerged as an efficient representation for photorealistic scene rendering. Instead of representing geometry with an explicit mesh or an implicit neural field, 3DGS models a scene as a collection of learnable 3D Gaussians, each associated with attributes such as position, scale, rotation, opacity, and color. These Gaussians are rasterized efficiently, enabling high-quality rendering at interactive frame rates. When adapted to dynamic human faces, this representation gives rise to \emph{Gaussian head avatars}, which model facial appearance and deformation using a set of animatable Gaussians. Compared with prior implicit representations, Gaussian head avatars offer a favorable balance between visual fidelity, rendering efficiency, and editability.

Based on the number of input views, Gaussian avatar reconstruction methods can be broadly divided into two categories. Multi-view methods based on calibrated camera arrays~\cite{xu2024gaussian,sun2025svg,teotia2024gaussianheads,aneja2025scaffoldavatar,giebenhain2024npga,qian2024gaussianavatars} achieve highly photorealistic reconstructions across diverse viewpoints and expressions, benefiting from dense observations of each subject. However, collecting such data requires specialized multi-camera capture systems, which greatly limits scalability and accessibility. To reduce this burden, recent works have explored monocular video-based methods~\cite{ma20243d,moon2025geoavatar,zhang2025hravatar,chen2024monogaussianavatar,wei2025gazegaussian,wei2025graphavatar,teotia2025audio,xiang2024flashavatar,song2025streamme,shao2024splattingavatar} and few-image-based methods~\cite{chu2024generalizable,zheng2025headgap,taubner2025cap4d,taubner2025mvp4d,kirschstein2025avat3r,he2025lam}. These methods incorporate additional priors, such as FLAME fitting~\cite{li2017learning} or semantic feature extraction~\cite{oquab2023dinov2}, to compensate for the lack of observations during training.

Nevertheless, Gaussian head avatars reconstructed from limited inputs remain fundamentally under-constrained. As a result, they often exhibit artifacts such as blurry textures, floating Gaussians, and unstable geometry in regions that are weakly observed during training. In contrast, multi-view Gaussian avatars typically provide cleaner and more stable renderings because their geometry and appearance are better constrained by dense observations. Our work is motivated by this gap: we aim to learn how degraded Gaussian avatar renderings should be refined using supervision derived from high-quality data, and to apply this knowledge to improve avatars reconstructed under sparse-view or monocular conditions.

\subsection{Generative Prior for Refinement}\label{sec:generative-prior}
Advances in generative models~\cite{goodfellow2014gan,ho2020denoising} have demonstrated not only impressive generation quality but also their utility for image refinement~\cite{saharia2022image,yang2021gan,saharia2022palette} and editing~\cite{rombach2021ldm,meng2021sdedit,lugmayr2022repaint}. This capability stems from expressive latent priors learned from large-scale data, which capture rich distributions of natural images. Such priors enable models both to generate photorealistic samples and to guide the refinement of degraded or incomplete inputs toward plausible, high-quality outputs. While most refinement methods have been developed for super-resolution~\cite{wang2024exploiting,wu2024seesr} and restoration~\cite{luo2024photo,zhang2024diff,wu2025difix3d+} in general scenes or images, only a few methods have been applied to human faces or avatars.

GAN-based methods typically encode images into the latent space of a pretrained model and then pass them through the generator again to produce modified images in 2D~\cite{richardson2021encoding,abdal2019image2stylegan,yun2024stylized,richardson2021psp} or 3D~\cite{yun2025ffacenerf,sun2022fenerf,jiang2022nerffaceediting}. These methods require images to be cropped and aligned. More closely related to our work, recent works~\cite{lee2025stylemm,wu2025text,wu2024portrait3d} utilize an SDEdit-like~\cite{meng2021sdedit} approach that adds noise and removes it to refine rendered avatars. While SDEdit-style refinement is effective for blind restoration, the network lacks explicit knowledge of how the avatar should look, and the stochastic noise-adding procedure introduces randomness. Consequently, the denoising step may produce inconsistent results across frames, leading to temporal jittering. 
Difix3D+~\cite{wu2025difix3d+} more directly couples a single-step diffusion refiner with optimization of static NeRF and 3DGS representations. EDGAR applies a related refinement strategy to animatable Gaussian head avatars, using paired avatar failures, a subject reference, and expression supervision to preserve identity and facial motion. Larger image-editing foundation models may offer higher single-frame quality, but avatar refinement also requires identity, expression, and cross-view consistency. EDGAR addresses these requirements through avatar-specific supervision and shared 3D optimization, without per-frame diffusion inference.

\section{\texorpdfstring{\revisedtext{Method}}{Method}}\label{sec:method}

Our goal is to design a model that refines rendered Gaussian avatars trained under diverse conditions, where limited observations often lead to visual artifacts. Specifically, we formulate avatar enhancement as a diffusion-prior-based refinement task: given a degraded rendering and a subject reference image, the model learns to produce a sharper and more faithful output. To enable this, we construct EDGAR-SET (Section~\ref{sec:edgar-set}), a triplet dataset that simulates realistic failure modes and provides paired ground-truth supervision. Building on this dataset, we train EDGAR, a diffusion-based refinement network conditioned on reference identity features, to correct artifacts while preserving subject fidelity (Section~\ref{sec:edgar}). Finally, we optimize the underlying Gaussian avatar itself using EDGAR-refined renderings as supervision, allowing the improvements to be absorbed into the 3D representation while preserving the efficiency of the original avatar model (Section~\ref{sec:refine}).

\begin{figure*}
  \includegraphics[width=\linewidth]{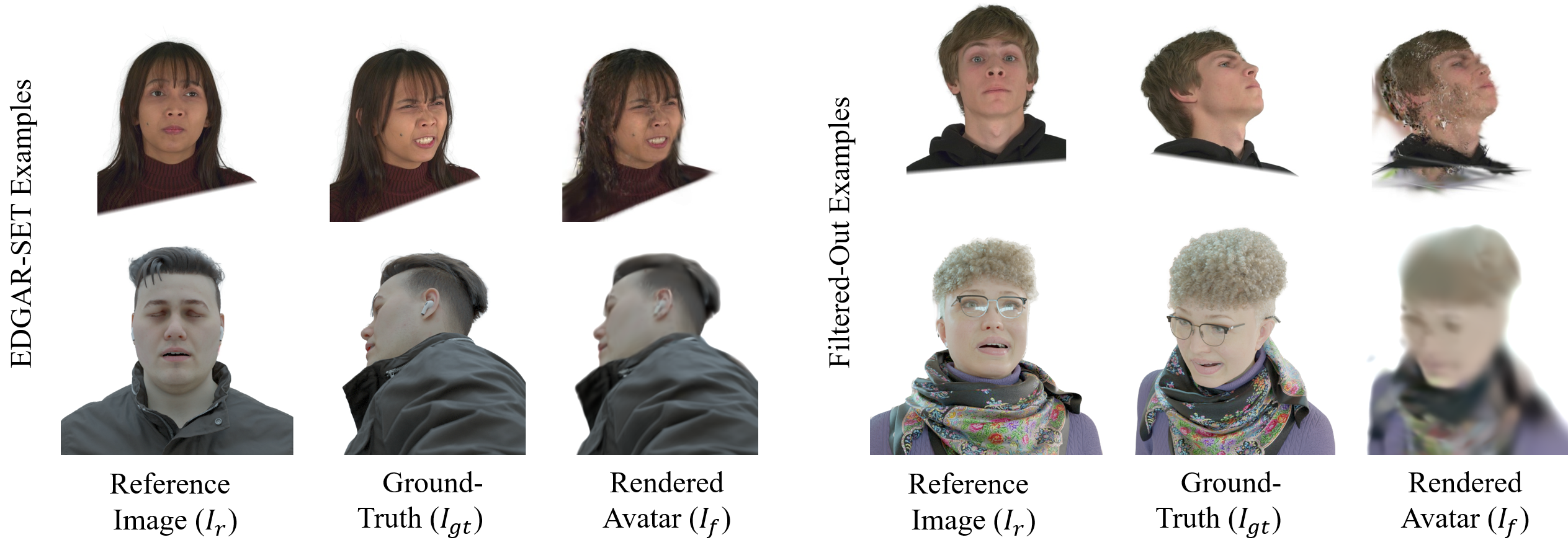}
  \caption{Examples from EDGAR-SET and discarded samples. We filter out cases where the geometry or expression is misaligned, making the recovery target ambiguous. Samples are removed when the combination of $I_f$ and $I_r$ from novel views or expressions cannot reliably reproduce $I_{gt}$.}
  \label{fig:edgarset-data}
\end{figure*}

\begin{figure*}\vspace{2mm}
  \includegraphics[width=0.84\linewidth]{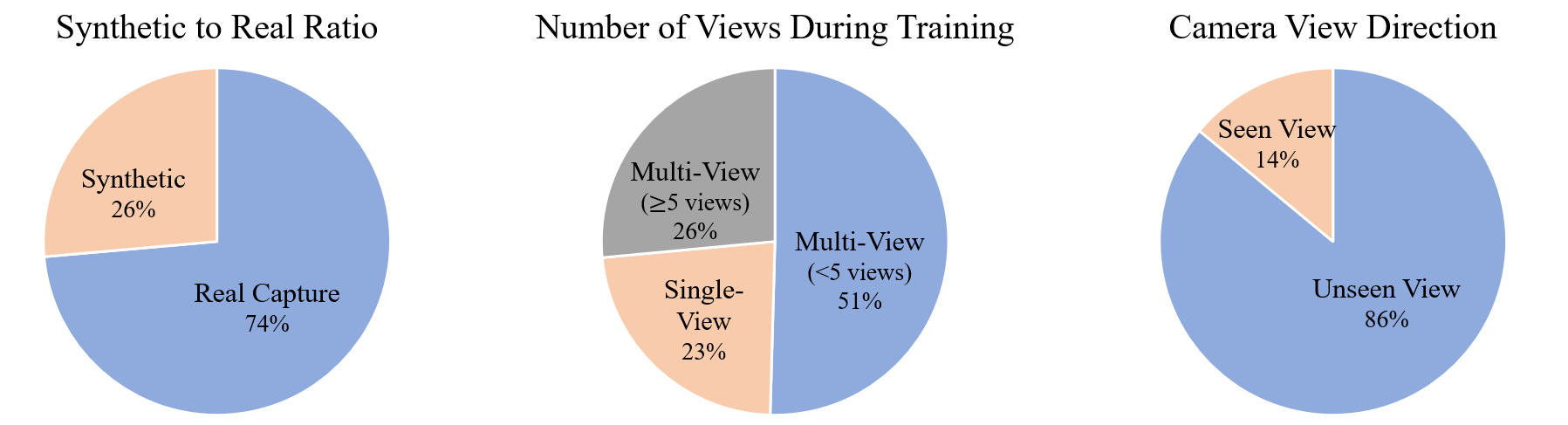}
  \caption{Configuration of EDGAR-SET. The dataset is constructed under diverse conditions to capture a wide range of Gaussian avatar failure cases. \revisedtext{The view-count groups (one, two to four, and at least five) are mutually exclusive and account for 23\%, 51\%, and 26\% of retained examples, respectively.}}
  \label{fig:edgarset-config}
\end{figure*}

\subsection{Synthesizing EDGAR-SET}\label{sec:edgar-set}
Training EDGAR requires paired data that captures how degraded Gaussian avatar renderings should be refined. To this end, we construct a triplet dataset, \textit{EDGAR-SET}, composed of synthetic failure images $I_f$, ground-truth images $I_{gt}$, and reference images $I_r$. The failure images $I_f$ simulate common artifacts observed in Gaussian avatars, while $I_{gt}$ provides the corresponding ground-truth image. The reference image $I_r$ is drawn from a ground-truth rendering of the same subject under a different view and/or expression, supplying appearance cues that guide refinement. At inference time, $I_r$ corresponds to a frame from the monocular video used to train the avatar. Given a degraded input $I_f$ and its reference $I_r$, EDGAR learns to generate a refined output $\hat{I}_{f}$ that faithfully approximates the ground truth $I_{gt}$.

To construct \textit{EDGAR-SET}, we define three key requirements to ensure effective training for avatar refinement. First, the dataset must contain diverse identities to support generalization across subjects. Second, it should include challenging conditions such as novel viewpoints, which are primary sources of artifacts in Gaussian avatar rendering. Third, each triplet must include a reference image that differs from the ground-truth target while still providing meaningful identity and appearance cues under varying poses or expressions. To satisfy these criteria, we design EDGAR-SET as a hybrid dataset that combines real-captured data with large-scale synthetic augmentation, enabling both realism and broad coverage of edge cases.

Following these guidelines, we construct EDGAR-SET using both real and synthetic data sources. For real data, we leverage the calibrated multi-view corpus NeRSemble~\cite{kirschstein2023nersemble} and build avatars using the RGBAvatar pipeline~\cite{li2025rgbavatar}, which provides high-fidelity reconstructions grounded in real-world appearance. For synthetic data, we incorporate the CAFCA~\cite{buehler2024cafca} corpus, which significantly expands identity diversity and introduces extreme view directions and expressions that are difficult to capture in real-world settings. Using CAFCA, we construct avatars based on the original 3D Gaussian Splatting formulation~\cite{kerbl20233d}, enabling the systematic simulation of a wide range of degradation patterns. The degradation patterns include floating Gaussian artifacts, blurry or over-smoothed regions, misaligned geometry, and view-dependent inconsistencies. The synthetic split primarily exhibits moderate failures due to a larger number of input views, while the real-captured split introduces more extreme failure modes, such as severe geometric distortions caused by limited-view reconstruction. Across both datasets, we vary training conditions—such as the number of input views—to induce realistic failure cases.

After constructing avatars, we obtain degraded renderings $I_f$ alongside their corresponding ground-truth images $I_{gt}$, and sample a distinct ground-truth frame as the reference $I_r$. This unified pipeline yields diverse triplets that cover both realistic and extreme scenarios, enabling EDGAR to learn robust refinement across a broad spectrum of avatar artifacts. In this way, our triplet formulation extends the utility of the original datasets beyond static ground-truth collections. Specifically, our triplet formulation converts the original datasets into supervised training data for refinement by simulating failures under diverse scenarios and incorporating reference images. In contrast, the original datasets (NeRSemble and CAFCA) primarily provide ground-truth observations without explicit degradation modeling.

From this raw triplet set, we filter out images with severe artifacts where $I_f$ diverges significantly from $I_{gt}$, making them unsuitable for training. We compute LPIPS scores as a similarity measure and discard cases with consistently low similarity. As shown in Figure~\ref{fig:edgarset-data}, triplets in which expression fidelity is lost or artifacts are irreparable are removed, while cases with degraded but recoverable quality—such as Gaussian noise artifacts with consistent expression and head pose—are retained. After this filtering step, the final \textit{EDGAR-SET}, consisting of 400K high-resolution images across thousands of identities, is constructed. Configuration of \textit{EDGAR-SET} is summarized in Figure~\ref{fig:edgarset-config}. \textit{EDGAR-SET} adheres to the guidelines established for training EDGAR, including diverse identities, expressions, and viewpoints, as well as the inclusion of reference images. \newtext{The resulting training regime emphasizes recoverable appearance artifacts and moderate reconstruction errors, whereas severe structural failures are not represented reliably by the available image and reference pairs.}

\subsection{EDGAR Model} \label{sec:edgar}
EDGAR is a conditional diffusion model that refines degraded inputs $I_f$ using a reference image $I_r$. For efficient inference, it is built upon the one-step diffusion model SD-Turbo~\cite{sauer2024adversarial}, a step-distilled version of \revisedtext{Stable Diffusion~\cite{rombach2021ldm}}, following prior work on efficient image refinement~\cite{wu2025difix3d+,yue2024efficient,parmar2024one}. Because our objective is to adapt a pretrained diffusion model to a Gaussian avatar refinement network, we freeze the residual blocks and cross-attention layers of SD-Turbo. The residual blocks are responsible for timestep injection, while the cross-attention layers handle prompt conditioning—neither of which we aim to modify. On the other hand, we update the self-attention layers, which govern pixel-to-pixel relationships, so that the network learns how to refine $I_f$. In addition, we pass the features from the encoder to a LoRA-added~\cite{hu2022lora} decoder with a zero-initialized residual layer so that the network better understands $I_f$, following ~\citet{parmar2024one}.

\begin{figure*} 
\centering
  \includegraphics[width=\linewidth]{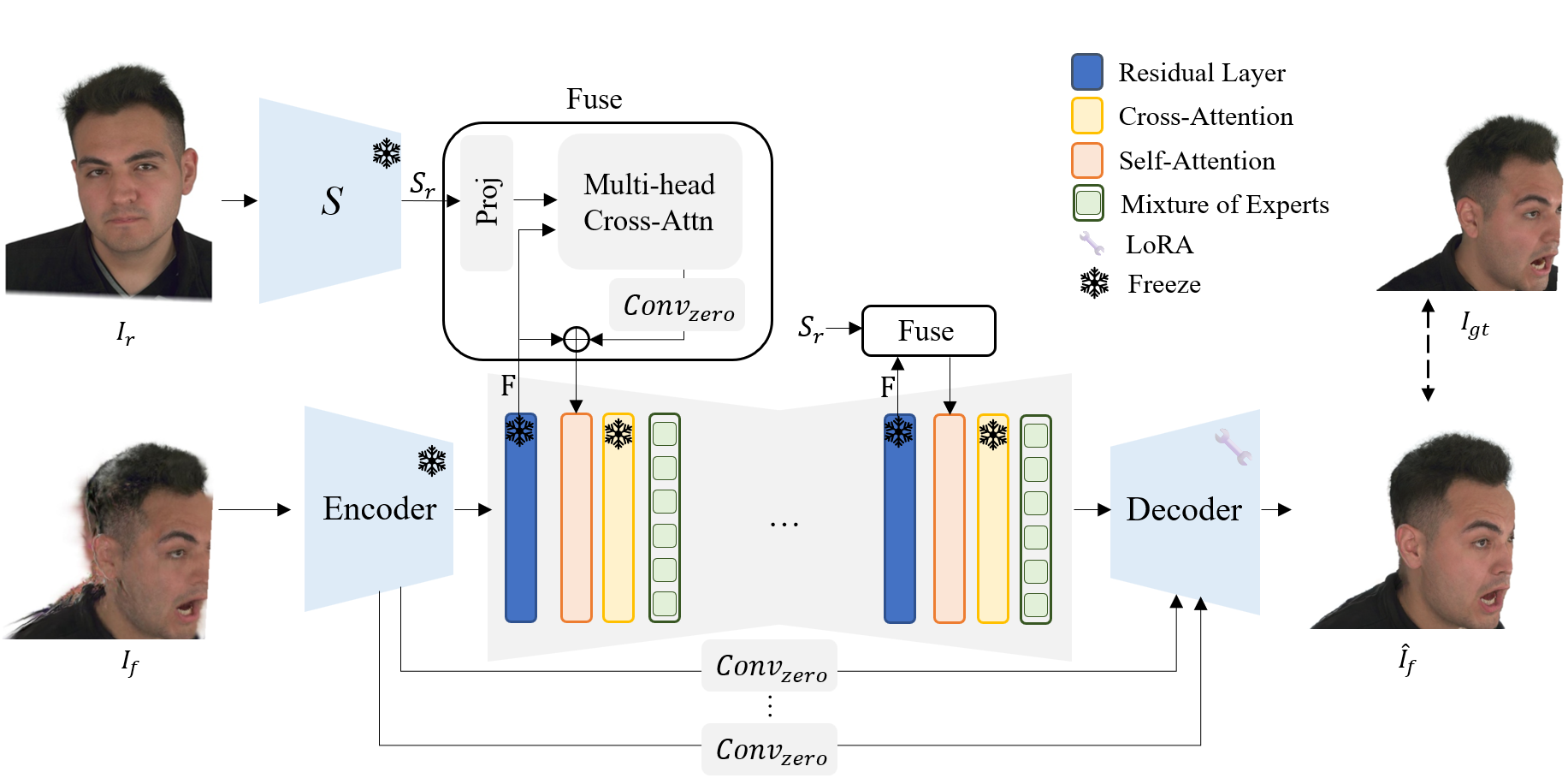} 
  \caption{Network architecture of EDGAR. EDGAR learns to refine rendered images of Gaussian avatars using EDGAR-SET. The fuse block in the Upblock of Diffusion UNet is simplified.}
  \label{fig:model}
\end{figure*}

To incorporate reference information, we implement a fusion block. Specifically, we pass $I_r$ through the pretrained Sapiens model~\cite{khirodkar2024sapiens}, a large self-supervised transformer known to capture rich human representations and perform well across diverse downstream tasks. We use Sapiens, denoted as $S$, as a feature extractor for reference identity. As illustrated in Figure~\ref{fig:model}, its features are projected and injected into the multi-head cross-attention layers to serve as keys and values, supplying identity-specific context that guides refinement, while the diffusion features $F$ from the previous layer serve as queries. This can be expressed as:

\begin{equation}
    \mathrm{Fuse}\big(F,S_r \big) 
    = \mathrm{Conv_\text{zero}} \big( \mathrm{softmax}\!\left(\frac{F\cdot \mathrm{Proj}(S_r)^\top}{\sqrt{d}}\right)\mathrm{Proj}(S_r) \big) + F,
\end{equation}
where  $S_r = S(I_r)$ denotes the Sapiens features from $I_r$, $\mathrm{Proj}$ is the projection layer, and $\mathrm{Conv_\text{zero}}$ is the zero-initialized convolutional layer. Note that because all added layers pass or are initialized with zeros including bias terms, in the beginning of the training, the model behaves exactly the same as the original SD-Turbo model. This initialization effectively preserves the performance of the pre-trained network while allowing it to smoothly converge to the refining network as training progresses.

To handle the wide range of degradation patterns encountered in Gaussian avatar refinement, we further adopt an MoE architecture. In practice, the input may exhibit substantially varying conditions, ranging from pronounced artifacts such as floating blobs and blurry Gaussians, to nearly clean reconstructions that require only minor corrections. The difficulty of refinement also varies depending on the reference condition, for example, whether the reference image is captured from a nearby or distant viewpoint. A single expert is often insufficient to model such diverse cases. We therefore introduce multiple experts that specialize in different refinement behaviors. Specifically, after the self-attention and cross-attention layers, a lightweight router predicts expert assignment scores from the intermediate feature and dispatches the feature to one selected expert for further processing. The entire routing mechanism is trained jointly with the refinement network in a self-supervised manner.

\subsection{Objectives} \label{sec:objectives}
To train the EDGAR model, we adopt image reconstruction losses for refined image $\hat{I}_f$ to follow the ground-truth image $I_{gt}$ conditioned on reference $I_r$. Specifically, we combine perceptual, pixel-level, and style consistency terms. The perceptual loss is measured using LPIPS~\cite{zhang2018lpips}, which encourages perceptual similarity between the refined output $\hat{I}_{f}$ and the ground-truth $I_{gt}$. The pixel loss is given by the $\ell_2$ distance, promoting accurate low-level reconstruction. In addition, we incorporate a Gram matrix loss computed from VGG features~\cite{gatys2016image}, which preserves higher-order correlations and encourages realistic texture synthesis. The total training objective is defined as follows:
\begin{equation} \label{eq:edgar_loss}
    \mathcal{L}_{img} = \| \hat{I}_{f} - I_{gt} \|_2^2 + \lambda_{\text{LPIPS}} \, \mathcal{L}_{\text{LPIPS}}(\hat{I}_{f}, I_{gt})
                + \lambda_{\text{Gram}} \, \mathcal{L}_{\text{Gram}}(\hat{I}_{f}, I_{gt}),
\end{equation}
where $\lambda_{\text{LPIPS}}$ and $\lambda_{\text{Gram}}$ control the relative weighting of each loss term.

In addition to the image-based loss $\mathcal{L}_{img}$, we introduce an expression consistency loss to better preserve subtle facial details such as the lips, eyes, and other fine-grained expression cues, which are often not sufficiently constrained by image reconstruction losses alone. To this end, we leverage the final-layer feature of POSTER V2~\cite{mao2025poster++}, a vision transformer-based facial expression model that jointly exploits landmark and image features for robust expression understanding. Rather than using its output logits, we use the pretrained feature representation as a perceptual expression descriptor, as intermediate features provide richer and smoother supervision for expression matching. Let $\phi(\cdot)$ denote the frozen POSTER V2 feature extractor. We then define the expression loss as
\begin{equation}
    \mathcal{L}_{\text{exp}} = \| \phi(\hat{I}_{f}) - \phi(I_{gt}) \|_2^2.
\end{equation}
This loss encourages the refined output to preserve the target facial expression more faithfully at a semantic feature level.

Finally, to prevent expert collapse, where only a small subset of experts are consistently selected, we employ a load-balancing loss following prior MoE formulations. Let $p_{i,e}$ denote the router probability of assigning sample $i$ to expert $e$, and let $z_i = \arg\max_e p_{i,e}$ be the selected expert index. We define the average routing importance and expert load as
\begin{equation}
\bar{p}_e = \frac{1}{N}\sum_{i=1}^{N} p_{i,e}, 
\qquad
\bar{l}_e = \frac{1}{N}\sum_{i=1}^{N} \mathbb{I}(z_i = e),
\end{equation}
where $N$ is the batch size and $\mathbb{I}(\cdot)$ is the indicator function. The load-balancing loss is then written as
\begin{equation}
\mathcal{L}_{\mathrm{lb}} = E \sum_{e=1}^{E} \bar{p}_e \bar{l}_e,
\end{equation}
where $E$ is the number of experts. This objective encourages both the router probabilities and the actual expert assignments to remain balanced across experts, improving specialization and stabilizing training.

In summary, the overall training objective for EDGAR is defined as
\begin{equation}~\label{eq:edgar}
    \mathcal{L}_{EDGAR} = \mathcal{L}_{img} + \lambda_{\text{exp}} \mathcal{L}_{\text{exp}} + \lambda_{\text{lb}} \mathcal{L}_{\text{lb}},
\end{equation}
where $\lambda_{\text{exp}}$ and $\lambda_{\text{lb}}$ control the contributions of the expression consistency loss and the load-balancing loss, respectively.

\begin{figure}[t]
  \includegraphics[width=\linewidth]{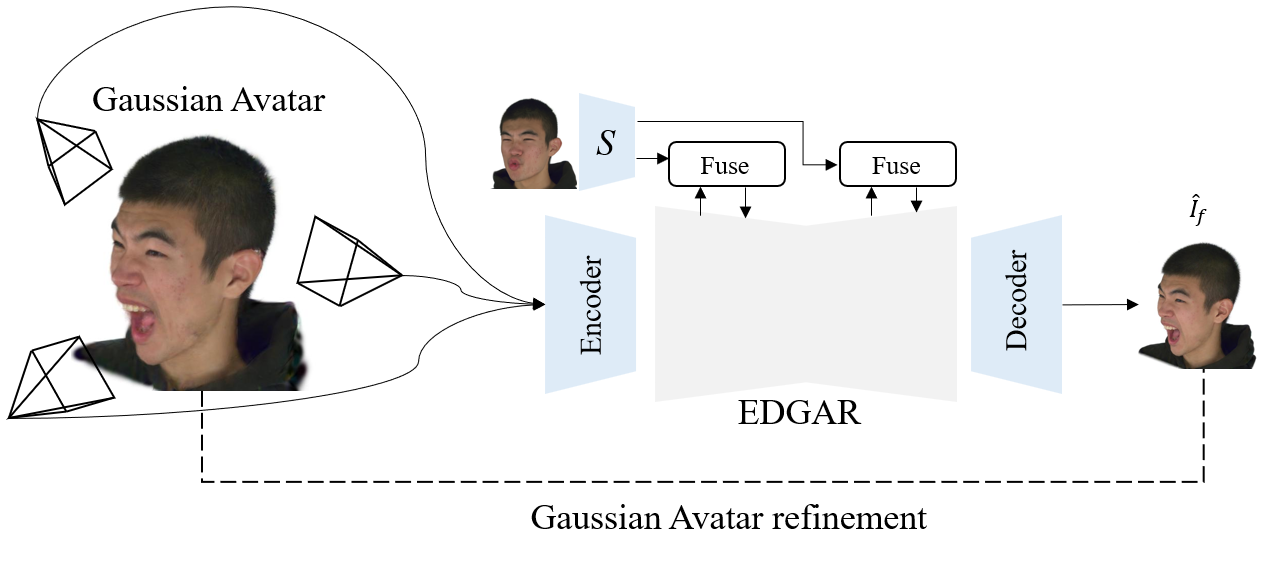}
  \caption{Avatar refinement process using EDGAR. Multi-view renderings of the source Gaussian avatar are refined by EDGAR and used as supervision targets for Gaussian avatar refinement.} \label{fig:inference}
\end{figure}

\subsection{Avatar Refinement}\label{sec:refine}
After training EDGAR, we use it to further refine the underlying Gaussian avatar itself. Given a Gaussian avatar as input, we first render it from multiple views and then apply EDGAR to each rendered image to obtain refined images $\hat{I}_{f}$. These refined images are subsequently used as supervision signals to optimize the Gaussian avatar parameters with an $\ell_1$ loss and an SSIM loss. Unlike prior refinement strategies that split or add Gaussians during optimization, we do not perform such operations, as they increase the number of Gaussians and lead to slower rendering. Our goal is to improve reconstruction quality while preserving the efficiency of the original model. During the refinement stage, a single reference image $I_r$ is randomly selected and used throughout the process. This refinement process is shown in Figure~\ref{fig:inference}.

Importantly, EDGAR is not used only as a post-processing module after rendering. Instead, its refined outputs are further used as optimization targets for the Gaussian avatar. In this way, the improvements introduced by EDGAR are absorbed into the 3D representation itself, allowing the refined avatar to produce higher-quality renderings directly. Because the refinement is performed by optimizing a single shared 3D Gaussian representation across multi-view rendered images, the resulting updates remain consistent in 3D and improve cross-view coherence.

Formally, letting $\mathcal{V}$ denote a set of sampled views, $R(\mathcal{G}, v)$ the rendering of Gaussian avatar $\mathcal{G}$ from view $v$, and $\hat{I}_{f}^{(v)}$ the EDGAR-refined result for that view, the avatar refinement objective is defined as follows:
\begin{equation} ~\label{eq:refine}
\revisedtext{\mathcal{L}_{\mathrm{avatar}} = \sum_{v \in \mathcal{V}} \left[ \lambda_{\mathrm{L1}} \| R(\mathcal{G}, v) - \hat{I}_{f}^{(v)} \|_1 + \lambda_{\mathrm{SSIM}} \, \mathcal{L}_{\mathrm{SSIM}}(R(\mathcal{G}, v), \hat{I}_{f}^{(v)}) \right],}
\end{equation} 
where $\lambda_{\mathrm{L1}}$ and $\lambda_{\mathrm{SSIM}}$ control the weights of the L1 and SSIM terms.

\begin{figure*}[t]
\centering
  \includegraphics[width=\linewidth]{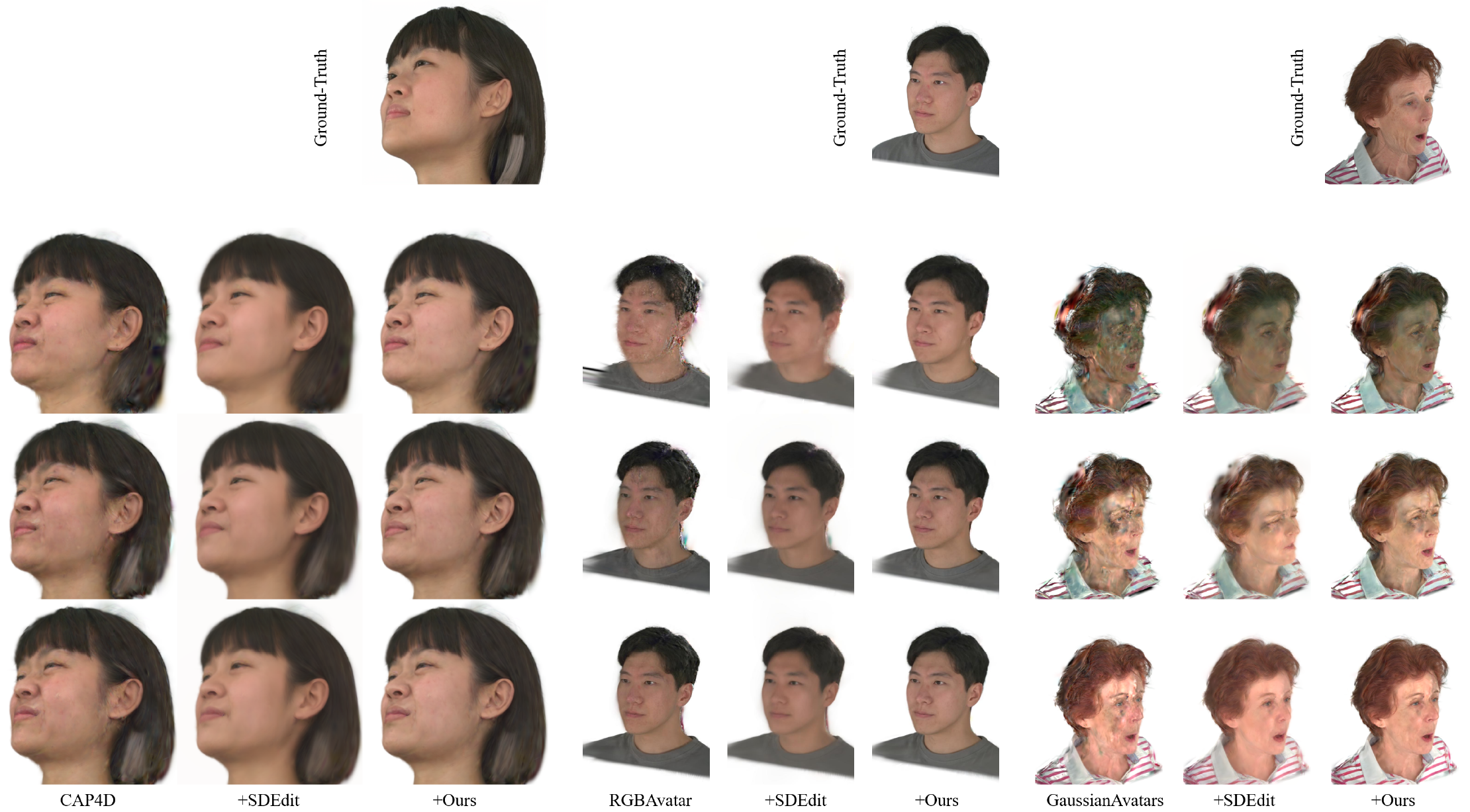} 
  \caption{Comparison of self-reenactment results. Our method refines each baseline by restoring facial details such as the eyes, wrinkles, and sharper hair, whereas SDEdit-based refinement alters the source identity. The upper rows show outputs from avatars trained or optimized with fewer input images or views (one image for CAP4D, one video for RGBAvatar, and eight videos for GaussianAvatars), while the lower rows show outputs from avatars trained with more images or videos. \newtext{Each block uses a different identity and input severity, so comparisons are made within each reconstruction method. The more severely degraded GaussianAvatars example is improved but remains only partially corrected.}}
  \label{fig:comp_self}
\end{figure*}

\section{Experiments} 

\subsection{Implementation Details}\label{sec:implementation}

We implemented and ran our data-construction pipeline using \revisedtext{RGBAvatar} and 3DGS on a computer equipped with a single NVIDIA RTX 3090 GPU. For the real-capture split of EDGAR-SET, we used \revisedtext{RGBAvatar} with the NeRSemble~\cite{kirschstein2023nersemble} dataset, varying the number of viewpoints, and performed self-reenactment to construct EDGAR-SET of 99 identities. Using 3DGS with CAFCA~\cite{buehler2024cafca}, we also constructed a synthetic split of EDGAR-SET of 1,500 identities with varying expressions, lighting conditions, and camera positions. While the real-capture split was constructed using 1 to 4 viewpoints during training, the synthetic split was constructed using only multi-view settings, specifically at least with 5 viewpoints, because 3DGS cannot reconstruct an avatar from a single viewpoint. All images were resized to 512×512 for both training and testing. During EDGAR-SET construction, LPIPS was used to compare the source and target images and approximately half of the pairs were filtered out.

We trained EDGAR on EDGAR-SET for 40,000 steps with a batch size of 4 and gradient accumulation over 2 iterations per step, which required approximately 40 hours. The learning rate was linearly increased up to $2\times 10^{-5}$ for 1,000 steps and constant afterward. The weights $\lambda_{LPIPS}$ and $\lambda_{Gram}$ used in Eq.~\eqref{eq:edgar_loss} were both set to 1.0, while the weights $\lambda_{exp}$ and $\lambda_{lb}$ used in Eq.~\eqref{eq:edgar} were set to 8.0 and 0.001. We set the timestep to $\tau = 250$ and used the prompt “High resolution portrait.” for both training and inference for text condition. A total of six expert MLPs were used in the MoE architecture. For the Gaussian avatar refinement process, we used the output Gaussians of each baseline model as input to EDGAR, rendered 50 views, and optimized the avatar for 500 steps. The weights $\lambda_{\mathrm{L1}}$ and $\lambda_{\mathrm{SSIM}}$ used in Eq.~\eqref{eq:refine} were set to 0.6 and 0.4, respectively. EDGAR was trained and evaluated on a single NVIDIA A6000 GPU, while the Gaussian avatar refinement stage was run on a single NVIDIA RTX 3090 GPU.

\paragraph{Computational cost.}
\newtext{Table~\ref{tab:cost_aux} reports the complete per-avatar refinement time, covering the full process from loading the avatar and EDGAR model to generating pseudo-targets, optimizing for 500 steps, and saving the checkpoint. Base-avatar optimization times are measured excluding the preprocessing and CAP4D's upstream view synthesis. Final-avatar rendering throughput is benchmarked separately.}

\begin{figure*}[!h]
\centering
  \includegraphics[width=0.95\linewidth]{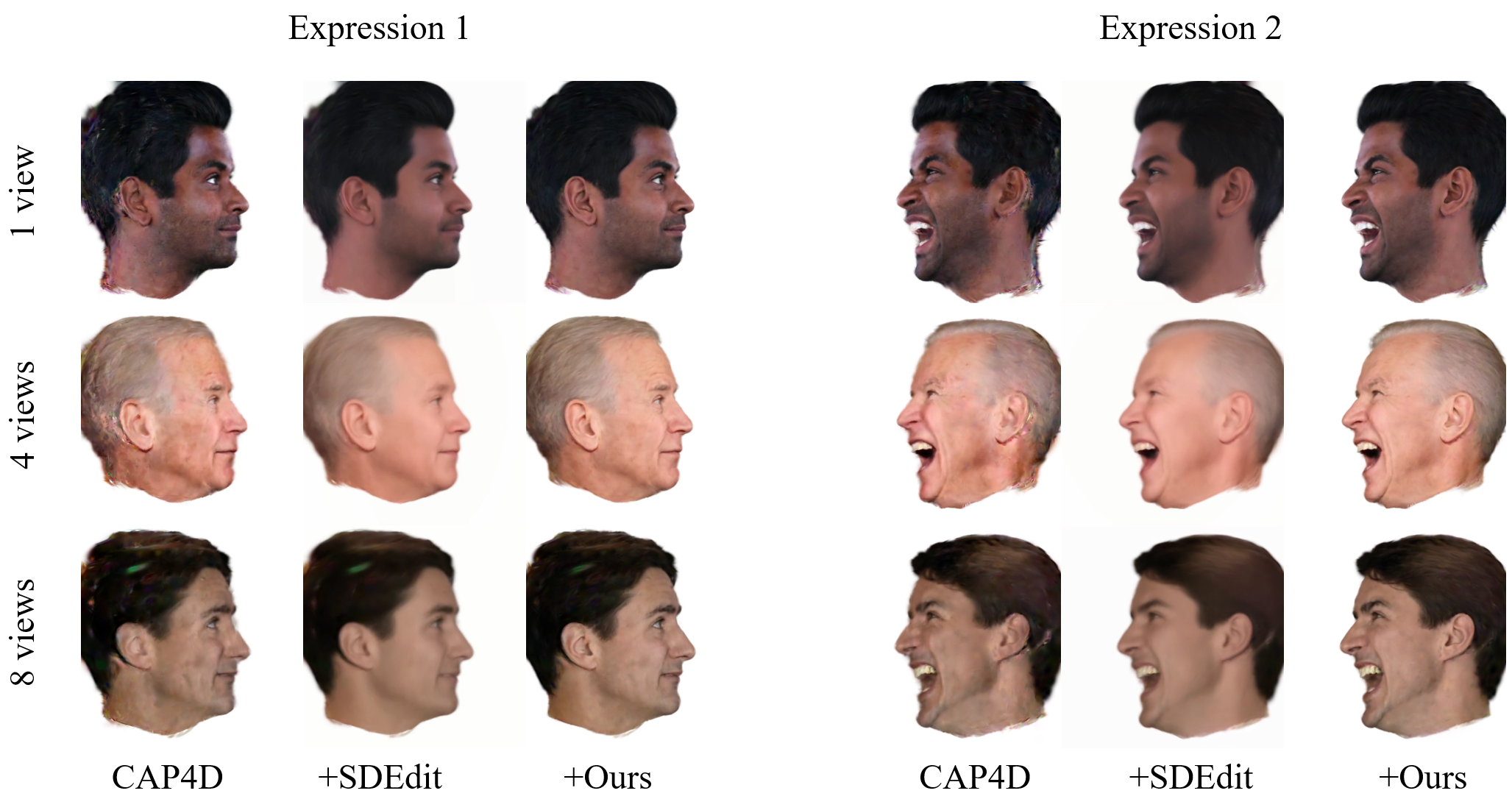}
  \caption{Comparison of cross-reenactment results rendered at novel views. Our method refines each face by restoring facial details while effectively removing Gaussian artifacts such as floating blobs. In contrast, SDEdit-based refinement introduces identity shifts and oversmoothing skin texture.}
  \label{fig:comp_cross}
\end{figure*}

\subsection{Baseline Comparison}\label{sec:comparison}

EDGAR is a general Gaussian avatar refinement framework and can
be applied to arbitrary trained or optimized Gaussian avatar reconstruction methods. To evaluate its effectiveness, we conducted experiments on three different baselines: CAP4D~\cite{taubner2025cap4d}, which is a single- to multi-image-based reconstruction method, GaussianAvatars~\cite{qian2024gaussianavatars}, which enables high-quality avatar reconstruction from multi-view videos, and RGBAvatar~\cite{li2025rgbavatar}, which is a single-video-based reconstruction method. These three methods represent complementary design choices, ranging from single-image-based and single-video-based to multi-view settings, providing a comprehensive benchmark for assessing EDGAR’s refinement capability across different reconstruction settings. For a comprehensive evaluation, because each method supports reconstruction from a different number of inputs, we conducted comparisons using avatars reconstructed with varying numbers of inputs. Specifically, we used 1, 4, and 8 images for CAP4D, 1, 4, and 8 videos for RGBAvatar, and 8, 10, and 12 videos for GaussianAvatars, following the settings used in each method. Note that RGBAvatar was used during data construction and is included here as an auxiliary test baseline. For testing, we used ten unseen identities from NeRSemble for self-reenactment and additionally six different identities from the INSTA~\cite{zielonka2023instant} dataset for cross-identity reenactment, and all comparisons were conducted on novel views.

In addition, following prior studies~\cite{wu2024portrait3d,wu2025text,lee2025stylemm}, we also applied an SDEdit-style refinement baseline to the generated avatars. Specifically, SD-Turbo was utilized, which is equivalent to EDGAR's backbone model with the timestep set to $\tau=250$ for noise injection and denoising. This enables a direct comparison between our refinement approach and diffusion-based editing methods widely used in 3D character generation.

\begin{figure*}[t]
\centering
\includegraphics[width=0.9\textwidth]{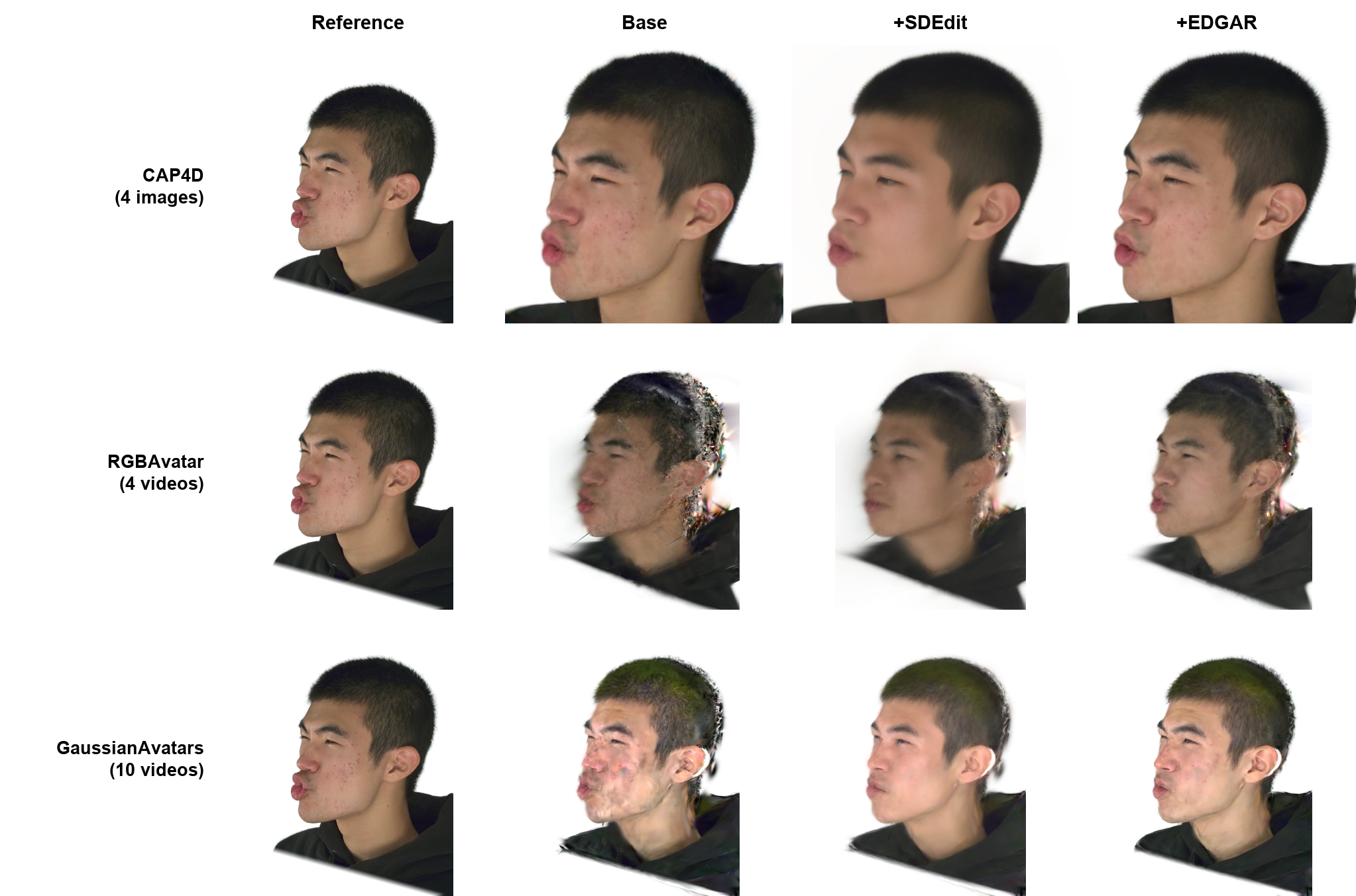}
\caption{\newtext{Matched-subject comparison at the same evaluation frame. Native inputs are four images for CAP4D, four videos for RGBAvatar, and ten videos for GaussianAvatars. Identity and evaluation frame are shared, while input regimes and initial reconstruction quality differ across pipelines.}}
\label{fig:matched_subject}
\end{figure*}

\revisedtext{Figure~\ref{fig:comp_self} presents EDGAR results for avatars reconstructed by CAP4D~\cite{taubner2025cap4d}, GaussianAvatars~\cite{qian2024gaussianavatars}, and RGBAvatar~\cite{li2025rgbavatar} under different input settings. Relative to the unrefined outputs, EDGAR reduces blurring and Gaussian artifacts while recovering fine facial details such as hair and skin texture.} In contrast, the SDEdit-style refinement often introduced stochastic variations, leading to oversmooth texture and identity drift. 
\revisedtext{These qualitative comparisons illustrate EDGAR's refinement behavior across Gaussian avatars generated by different reconstruction methods.} \newtext{To further demonstrate performance on the same identity, Figure~\ref{fig:matched_subject} complements Figure~\ref{fig:comp_self} by presenting the results of the three pipelines on the same identity and frame.}

This observation is further supported by the quantitative results for self-reenactment reported in Tables~\ref{tab:cap4d}, \ref{tab:rgba}, and \ref{tab:gsa}. \revisedtext{On CAP4D, EDGAR obtains the highest CLIP similarity and lowest $\ell_1$ error at every input count, while the Baseline retains the lowest LPIPS and SDEdit the highest PSNR. On RGBAvatar and GaussianAvatars, EDGAR leads all four metrics at every input count.}

\revisedtext{We also evaluated cross-identity reenactment.} Because ground-truth images are unavailable in this setting, we instead used cosine similarity between identity embeddings (CSIM), following CAP4D~\cite{taubner2025cap4d} for evaluation.  As shown in Table~\ref{tab:cross}, although EDGAR did not surpass the input avatar in identity similarity, the degradation was negligible. In contrast, SDEdit led to a substantial drop in CSIM scores. This trend is also reflected in Figure~\ref{fig:comp_cross}. EDGAR effectively removed artifacts around the ear, neck, and hair present in the base model, while SDEdit introduced identity changes and excessive smoothing.

\newcommand{\best}[1]{\cellcolor{red!25}{#1}}
\newcommand{\second}[1]{\cellcolor{yellow!25}{#1}}

\begin{table*}[t]
\centering
\caption{Quantitative comparison on CAP4D across different numbers of input views. Best results per column are in red and second best are in yellow.}
\renewcommand{\arraystretch}{1.1}
\resizebox{0.98\textwidth}{!}{
\begin{tabular}{l|cccc|cccc|cccc}
\hline
\multirow{2}{*}{Method}
& \multicolumn{4}{c|}{1 view}
& \multicolumn{4}{c|}{4 views}
& \multicolumn{4}{c}{8 views} \\
& CLIP $\uparrow$ & L1 $\downarrow$ & LPIPS $\downarrow$ & PSNR $\uparrow$
& CLIP $\uparrow$ & L1 $\downarrow$ & LPIPS $\downarrow$ & PSNR $\uparrow$
& CLIP $\uparrow$ & L1 $\downarrow$ & LPIPS $\downarrow$ & PSNR $\uparrow$ \\
\hline
CAP4D
& \second{0.9369} & 0.0584 & \best{0.2133} & 17.5651
& \second{0.9393} & 0.0537 & \best{0.1931} & 18.1435
& \second{0.9462} & \second{0.0518} & \best{0.1847} & 18.4773 \\

+SDEdit
& 0.9078 & \second{0.0581} & 0.2728 & \best{17.7737}
& 0.9101 & \second{0.0535} & 0.2562 & \best{18.3674}
& 0.9151 & \second{0.0518} & 0.2529 & \best{18.6645} \\

+Ours
& \best{0.9449} & \best{0.0580} & \second{0.2147} & \second{17.5978}
& \best{0.9488} & \best{0.0534} & \second{0.1962} & \second{18.1927}
& \best{0.9531} & \best{0.0513} & \second{0.1884} & \second{18.5329} \\
\hline
\end{tabular}
}
\label{tab:cap4d}
\end{table*}

\begin{table*}[t]
\centering
\caption{Quantitative comparison on RGBAvatar across different numbers of input \revisedtext{videos}. Best results per column are in red and second best are in yellow.}
\renewcommand{\arraystretch}{1.1}
\resizebox{\textwidth}{!}{
\begin{tabular}{l|cccc|cccc|cccc}
\hline
\multirow{2}{*}{Method}
& \multicolumn{4}{c|}{1 \revisedtext{video}}
& \multicolumn{4}{c|}{4 \revisedtext{videos}}
& \multicolumn{4}{c}{8 \revisedtext{videos}} \\
& CLIP $\uparrow$ & L1 $\downarrow$ & LPIPS $\downarrow$ & PSNR $\uparrow$
& CLIP $\uparrow$ & L1 $\downarrow$ & LPIPS $\downarrow$ & PSNR $\uparrow$
& CLIP $\uparrow$ & L1 $\downarrow$ & LPIPS $\downarrow$ & PSNR $\uparrow$ \\
\hline
RGBAvatar
& 0.8423 & 0.0832 & 0.3619 & 16.0469
& \second{0.8877} & \second{0.0552} & \second{0.2524} & 19.4950
& \second{0.9210} & \second{0.0502} & \second{0.2100} & \second{20.5128} \\

+SDEdit
& \second{0.8554} & \second{0.0827} & \second{0.3601} & \second{16.5024}
& 0.8639 & 0.0570 & 0.2878 & \second{19.9181}
& 0.8712 & 0.0555 & 0.2752 & 20.5109 \\

+Ours
& \best{0.8897} & \best{0.0677} & \best{0.3096} & \best{17.1533}
& \best{0.9273} & \best{0.0364} & \best{0.2121} & \best{21.5199}
& \best{0.9379} & \best{0.0332} & \best{0.1894} & \best{22.9937} \\
\hline
\end{tabular}
}
\label{tab:rgba}
\end{table*}

\begin{table*}[t]
\centering
\caption{Quantitative comparison on GaussianAvatars across different numbers of input \revisedtext{videos}. Best results per column are in red and second best are in yellow.}
\renewcommand{\arraystretch}{1.1}
\resizebox{\textwidth}{!}{
\begin{tabular}{l|cccc|cccc|cccc}
\hline
\multirow{2}{*}{Method}
& \multicolumn{4}{c|}{8 \revisedtext{videos}}
& \multicolumn{4}{c|}{10 \revisedtext{videos}}
& \multicolumn{4}{c}{12 \revisedtext{videos}} \\
& CLIP $\uparrow$ & L1 $\downarrow$ & LPIPS $\downarrow$ & PSNR $\uparrow$
& CLIP $\uparrow$ & L1 $\downarrow$ & LPIPS $\downarrow$ & PSNR $\uparrow$
& CLIP $\uparrow$ & L1 $\downarrow$ & LPIPS $\downarrow$ & PSNR $\uparrow$ \\
\hline
GaussianAvatars
& 0.7819 & 0.0852 & 0.3299 & 16.2855
& 0.8407 & \second{0.0644} & 0.2661 & 18.5994
& 0.8759 & \second{0.0738} & 0.2303 & \second{18.1037} \\

+SDEdit
& \second{0.8173} & \second{0.0806} & \second{0.2894} & \second{16.9214}
& \second{0.8767} & 0.0649 & \second{0.2377} & \second{18.9219}
& \second{0.9025} & 0.0782 & \second{0.2202} & 18.0437 \\

+Ours
& \best{0.8300} & \best{0.0739} & \best{0.2526} & \best{17.3766}
& \best{0.8977} & \best{0.0593} & \best{0.2050} & \best{19.3206}
& \best{0.9214} & \best{0.0724} & \best{0.1875} & \best{18.2789} \\
\hline
\end{tabular}
}
\label{tab:gsa}
\end{table*}

\begin{table}[t]
\centering
\caption{Quantitative comparison of CSIM across different numbers of input views. Best results per column are in red and second best are in yellow.}
\resizebox{0.7\linewidth}{!}{
\begin{tabular}{l|c|c|c}
\hline
{Method}
& 1 view
& 4 views
& 8 views \\
\hline
Base
& \best{0.5510}
& \best{0.5642}
& \best{0.5651} \\
+SDEdit
& 0.3178
& 0.3432
& 0.3380 \\
+Ours
& \second{0.5169}
& \second{0.5326}
& \second{0.5419} \\
\hline
\end{tabular}
}
\label{tab:cross}
\end{table}

\begin{table*}[t]
\centering
\caption{Ablation study results across different numbers of input views on self-reenactment. Best results per column are in red and second best are in yellow.}
\renewcommand{\arraystretch}{1.1}
\resizebox{\textwidth}{!}{
\begin{tabular}{l|cccc|cccc|cccc}
\hline
\multirow{2}{*}{Method}
& \multicolumn{4}{c|}{1 view}
& \multicolumn{4}{c|}{4 views}
& \multicolumn{4}{c}{8 views} \\
& CLIP $\uparrow$ & L1 $\downarrow$ & LPIPS $\downarrow$ & PSNR $\uparrow$
& CLIP $\uparrow$ & L1 $\downarrow$ & LPIPS $\downarrow$ & PSNR $\uparrow$
& CLIP $\uparrow$ & L1 $\downarrow$ & LPIPS $\downarrow$ & PSNR $\uparrow$ \\
\hline

w/o $I_r$
& 0.9163 & 0.0595 & 0.2683 & 17.5822
& 0.9222 & 0.0547 & 0.2492 & 18.1634
& 0.9262 & 0.0527 & 0.2448 & 18.4787 \\

w/o Synthetic
& 0.9351 & 0.0583 & 0.2292 & \second{17.6128}
& 0.9407 & 0.0538 & 0.2106 & 18.1925
& 0.9436 & 0.0518 & 0.2033 & 18.5236 \\

w/o $\mathcal{L}_{exp}$
& 0.9373 & \best{0.0578} & 0.2297 & \best{17.6271}
& 0.9433 & \best{0.0533} & 0.2107 & \best{18.2135}
& 0.9465 & \best{0.0512} & 0.2038 & \best{18.5370} \\

w/o MoE
& 0.9420 & \second{0.0580} & \second{0.2148} & 17.5757
& 0.9456 & \second{0.0534} & \best{0.1962} & 18.1689
& 0.9503 & 0.0514 & \best{0.1884} & 18.5053 \\

w/ two Experts
& \second{0.9433} & 0.0581 & 0.2154 & 17.5824
& \second{0.9470} & 0.0535 & \second{0.1967} & 18.1802
& \second{0.9521} & 0.0514 & \second{0.1888} & 18.5201 \\

Ours
& \best{0.9449} & \second{0.0580} & \best{0.2147} & 17.5978
& \best{0.9488} & \second{0.0534} & \best{0.1962} & \second{18.1927}
& \best{0.9531} & \second{0.0513} & \best{0.1884} & \second{18.5329} \\

\hline
\end{tabular}
}
\label{tab:ablation}
\end{table*}

\begin{table}[t]
\centering
\caption{Ablation study results across different numbers of input views on cross-reenactment measured by CSIM. Best results per column are in red and second best are in yellow.}
\resizebox{0.85\linewidth}{!}{
\begin{tabular}{l|c|c|c}
\hline
{Method}
& 1 view
& 4 views
& 8 views \\
\hline

w/o $I_r$
& 0.4508
& 0.4716
& 0.4769 \\
w/o Synthetic
& 0.4971
& 0.5244
& 0.5242 \\
w/o $\mathcal{L}_{exp}$
& 0.4645
& 0.4930
& 0.4927 \\
w/o MoE
& \best{0.5177}
& 0.5287
& 0.5355 \\
w/ two Experts
& 0.5162
& \second{0.5301}
& \second{0.5363} \\
Ours
& \second{0.5169}
& \best{0.5326}
& \best{0.5419} \\
\hline
\end{tabular}
}
\label{tab:cross-ablation}
\end{table}

\subsection{\newtext{Temporal Evaluation and Cost Analysis}}\label{sec:auxiliary}

\newtext{We measured foreground temporal warping error (tWE) on 100 consecutive frames for each NeRSemble test identity. Dense optical flow and a bidirectional cycle-consistency validity mask were estimated from each unrefined Base sequence, where the same correspondences were used for Base and EDGAR. We computed the mean RGB $L_1$ difference after warping on valid foreground pixels at $256\times256$ resolution and averaged the errors over the 99 transitions within each identity. We report the mean and standard deviation across identities. As presented in Table~\ref{tab:temporal_aux}, EDGAR lowered tWE for CAP4D and RGBAvatar but raised it for GaussianAvatars, although the average tWE across pipelines decreased. This metric measures motion-compensated photometric stability, complementing the spatial-fidelity evaluations.}

\begin{table}[t]
\centering
\caption{\newtext{Foreground tWE ($\times10^{-3}$; lower is better). Values are mean $\pm$ identity-level standard deviation. The lower value in each row is red.}}
\resizebox{0.8\columnwidth}{!}{%
{\newmaterialcolor
\begin{tabular}{lcc}
\toprule
Pipeline & Base & +EDGAR \\
\midrule
CAP4D & {$1.759\pm0.210$} & \best{$1.696\pm0.199$} \\
RGBAvatar & {$7.424\pm1.411$} & \best{$5.574\pm1.512$} \\
GaussianAvatars & \best{$4.929\pm1.111$} & {$5.938\pm1.932$} \\
Average & {$4.704\pm0.803$} & \best{$4.403\pm1.105$} \\
\bottomrule
\end{tabular}}}
\label{tab:temporal_aux}
\end{table}

\begin{table*}[t]
\centering
\caption{\newtext{Computational cost and output throughput on an RTX 3090. Refinement time includes loading, pseudo-target generation, avatar optimization, and checkpoint writing. Values are mean $\pm$ sample standard deviation.}}
\resizebox{0.65\textwidth}{!}{%
{\newmaterialcolor
\begin{tabular}{lrrrr}
\toprule
\multirow{2}{*}{Pipeline} & \multicolumn{2}{c}{One-time cost (min)} & \multicolumn{2}{c}{Output FPS} \\
\cmidrule(lr){2-3}\cmidrule(lr){4-5}
& Base opt. & Avatar refine. & Base & EDGAR \\
\midrule
CAP4D & $484.0\pm80.5$ & $5.04\pm0.11$ & $50.3\pm0.3$ & $49.4\pm0.6$ \\
RGBAvatar & $105.6\pm4.4$ & $7.96\pm0.98$ & $813.8\pm36.4$ & $826.3\pm16.9$ \\
GaussianAvatars & $238.7\pm8.8$ & $16.60\pm5.98$ & $153.4\pm7.4$ & $156.1\pm0.7$ \\
\bottomrule
\end{tabular}}}
\label{tab:cost_aux}
\end{table*}

\newtext{For the cost analysis, we estimated Base-avatar optimization times for CAP4D, RGBAvatar, and GaussianAvatars and measured the complete EDGAR refinement time. Base reconstruction times were based on 100,000 CAP4D updates, 60,000 RGBAvatar updates, and 750,000 GaussianAvatars updates. They exclude preprocessing and CAP4D's separate upstream multi-view synthesis stage, reported to require approximately four hours~\cite{taubner2025cap4d}. As presented in Table~\ref{tab:cost_aux}, complete EDGAR refinement required 5.04, 7.96, and 16.60 minutes on average, respectively. Importantly, EDGAR incurs only a one-time refinement cost and adds no inference-time overhead: the refined avatars preserve the original Gaussian count and rendering path. In our experiments, output throughput remained within 2\% of the base models and increased slightly in two of the three pipelines.}

\subsection{\newtext{Cross-View Geometric Consistency}}
\label{sec:cross-view-consistency}

\newtext{We additionally measured whether refinement changed the 3D consistency of the reconstructed avatar across viewpoints. For each identity and input setting, we sampled 15 evenly spaced time steps and evaluated four views, averaging both directions of every view pair. For GaussianAvatars and RGBAvatar, these were synchronized cameras. Because CAP4D's stored sequence changes facial state together with viewpoint, we instead rendered each fixed facial state from four nearby novel cameras. Frame-level values were first averaged within each identity, and Table~\ref{tab:cross-view-consistency} reports the mean and sample standard deviation over the nine identities.}

\newtext{With calibrated renderer depth for CAP4D and GaussianAvatars, we unprojected a source pixel, transformed it into the target camera, and measured the symmetric relative depth discrepancy $E_d=2|z-d|/(|z|+|d|)$, where $z$ is its projected target-camera depth and $d$ is the sampled target depth. We also report valid reprojection coverage $C$, defined as the proportion of source-foreground pixels whose projections are valid in the target view. RGBAvatar does not expose comparable renderer depth for its stored views, so we computed its $E_d$ and $C$ values using cameras and depth estimated jointly from the four images by VGGT~\cite{wang2025vggt}. For calibrated CAP4D and GaussianAvatars, the refined avatar improved both $E_d$ and $C$ at every input setting. For RGBAvatar, whose geometry was estimated from the rendered images, refinement improved both measures at four and eight inputs but worsened both in the single-input setting. Overall, these results show that EDGAR generally improves the 3D cross-view consistency of the refined avatars.}


\begin{table}[!t]
\centering
\caption{Cross-view geometric consistency results. Values are reported as mean ± standard deviation. Better values are highlighted in red.}
\label{tab:cross-view-consistency}
\scriptsize
\setlength{\tabcolsep}{2.4pt}
\renewcommand{\arraystretch}{1.05}
\resizebox{\columnwidth}{!}{%
\begin{tabular}{llcrr}
\toprule
Avatar family & Inputs & Metric & Base & EDGAR \\
\midrule
\multirow{6}{*}{CAP4D}
 & \multirow{2}{*}{1}
 & $E_d\downarrow$ & {$0.0028\pm0.0004$} & \best{$0.0025\pm0.0003$} \\
 & & $C\uparrow$ & {$0.9135\pm0.0181$} & \best{$0.9217\pm0.0159$} \\
\cmidrule(lr){2-5}
 & \multirow{2}{*}{4}
 & $E_d\downarrow$ & {$0.0028\pm0.0004$} & \best{$0.0025\pm0.0004$} \\
 & & $C\uparrow$ & {$0.9145\pm0.0166$} & \best{$0.9222\pm0.0153$} \\
\cmidrule(lr){2-5}
 & \multirow{2}{*}{8}
 & $E_d\downarrow$ & {$0.0028\pm0.0004$} & \best{$0.0025\pm0.0004$} \\
 & & $C\uparrow$ & {$0.9141\pm0.0169$} & \best{$0.9218\pm0.0157$} \\
\midrule
\multirow{6}{*}{RGBAvatar}
 & \multirow{2}{*}{1}
 & $E_d\downarrow$ & \best{$0.0083\pm0.0018$} & {$0.0097\pm0.0037$} \\
 & & $C\uparrow$ & \best{$0.4855\pm0.0453$} & {$0.4261\pm0.0753$} \\
\cmidrule(lr){2-5}
 & \multirow{2}{*}{4}
 & $E_d\downarrow$ & {$0.0088\pm0.0020$} & \best{$0.0074\pm0.0017$} \\
 & & $C\uparrow$ & {$0.4248\pm0.0445$} & \best{$0.4339\pm0.0383$} \\
\cmidrule(lr){2-5}
 & \multirow{2}{*}{8}
 & $E_d\downarrow$ & {$0.0094\pm0.0021$} & \best{$0.0086\pm0.0034$} \\
 & & $C\uparrow$ & {$0.4233\pm0.0372$} & \best{$0.4348\pm0.0371$} \\
\midrule
\multirow{6}{*}{GaussianAvatars}
 & \multirow{2}{*}{8}
 & $E_d\downarrow$ & {$0.0070\pm0.0008$} & \best{$0.0063\pm0.0006$} \\
 & & $C\uparrow$ & {$0.5168\pm0.0418$} & \best{$0.5336\pm0.0420$} \\
\cmidrule(lr){2-5}
 & \multirow{2}{*}{10}
 & $E_d\downarrow$ & {$0.0072\pm0.0009$} & \best{$0.0064\pm0.0007$} \\
 & & $C\uparrow$ & {$0.5112\pm0.0402$} & \best{$0.5261\pm0.0405$} \\
\cmidrule(lr){2-5}
 & \multirow{2}{*}{12}
 & $E_d\downarrow$ & {$0.0069\pm0.0010$} & \best{$0.0062\pm0.0007$} \\
 & & $C\uparrow$ & {$0.5131\pm0.0427$} & \best{$0.5265\pm0.0404$} \\
\bottomrule
\end{tabular}%
}
\end{table}



\subsection{\newtext{Pose-and-Expression Robustness}}\label{sec:robustness}

\newtext{We evaluated large-pose and open-mouth conditions, using four images for CAP4D, four videos for RGBAvatar, and ten videos for GaussianAvatars. For the video pipelines, we sampled 64 uniformly spaced time steps from each camera sequence. For CAP4D, we evaluated eight images per identity. We used landmarks estimated from ground-truth images using Face Mesh~\cite{kartynnik2019facial} to determine absolute head yaw through a perspective pose fit and mouth opening through the 3D inner-lip gap divided by mouth width. We report results for three subsets: $|\mathrm{yaw}|\geq30^\circ$, normalized lip gap $\geq0.10$, and their intersection. Subset membership was determined independently of the Base and EDGAR outputs. We measured foreground $L_1$ and LPIPS with a common foreground mask and white background. Images retained their aspect ratio, with the longest side resized to 512 pixels. Native ground-truth masks were used for the video pipelines, whereas CAP4D foreground was identified from its white background. Errors were averaged within each identity and then across identities (Table~\ref{tab:pose-expression}).}

\begin{table}[t]
\centering
\caption{\newtext{Pose-and-expression evaluation at representative native input settings (CAP4D: four images; RGBAvatar: four videos; GaussianAvatars: ten videos). Values are equally weighted identity means. Large pose denotes $|\mathrm{yaw}|\geq30^\circ$, and open mouth denotes a normalized lip gap $\geq0.10$. Better values are highlighted in red.}}\vspace{-2mm}
\resizebox{\columnwidth}{!}{%
{\newmaterialcolor
\setlength{\tabcolsep}{3pt}
\begin{tabular}{llcccc}
\toprule
& & \multicolumn{2}{c}{Foreground $L_1$$\downarrow$} & \multicolumn{2}{c}{Foreground LPIPS$\downarrow$} \\
\cmidrule(lr){3-4}\cmidrule(lr){5-6}
Pipeline & Condition & Base & EDGAR & Base & EDGAR \\
\midrule
CAP4D & Large pose & \best{0.07500} & 0.07601 & 0.22353 & \best{0.21704} \\
 & Open mouth & 0.06906 & \best{0.06660} & \best{0.21057} & 0.21975 \\
 & Both & \best{0.07475} & 0.08084 & \best{0.23054} & 0.23507 \\
\midrule
RGBAvatar & Large pose & 0.08894 & \best{0.06121} & 0.17845 & \best{0.14399} \\
 & Open mouth & 0.06410 & \best{0.05372} & 0.12051 & \best{0.10368} \\
 & Both & 0.08785 & \best{0.06112} & 0.17563 & \best{0.14150} \\
\midrule
GaussianAvatars & Large pose & 0.11531 & \best{0.10845} & 0.18029 & \best{0.13128} \\
 & Open mouth & 0.11748 & \best{0.11192} & 0.17604 & \best{0.12943} \\
 & Both & 0.11762 & \best{0.11187} & 0.17572 & \best{0.12934} \\
\bottomrule
\end{tabular}}}
\label{tab:pose-expression}
\end{table}

\begin{figure}[t]
\centering
\setlength{\tabcolsep}{1.5pt}
\begin{tabular}{ccc}
\newtext{Ground truth} & \newtext{Base} & \newtext{EDGAR} \\
\includegraphics[width=.31\columnwidth]{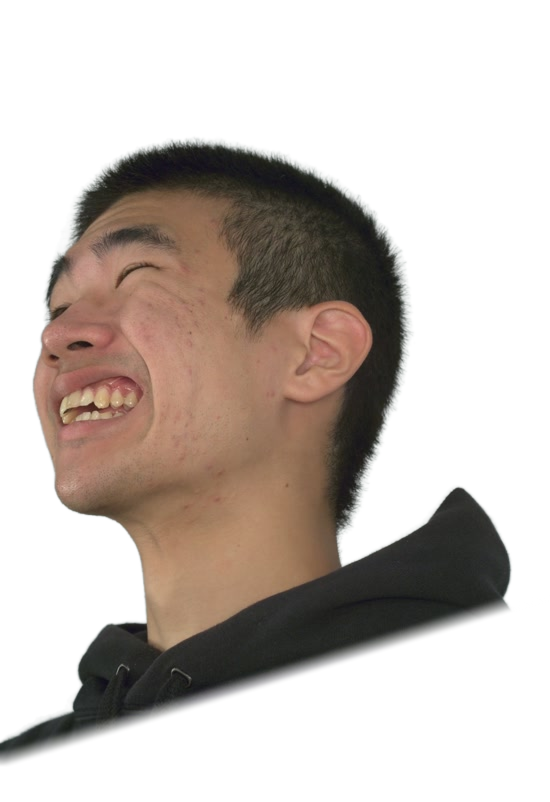} &
\includegraphics[width=.31\columnwidth]{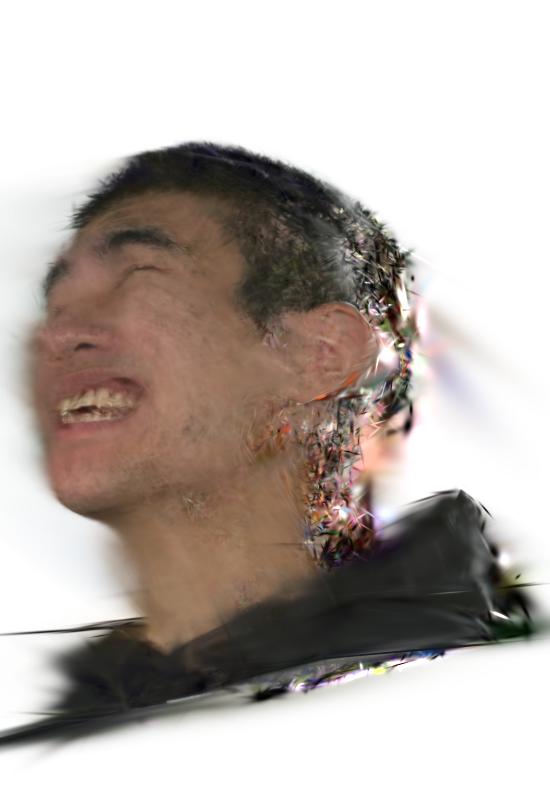} &
\includegraphics[width=.31\columnwidth]{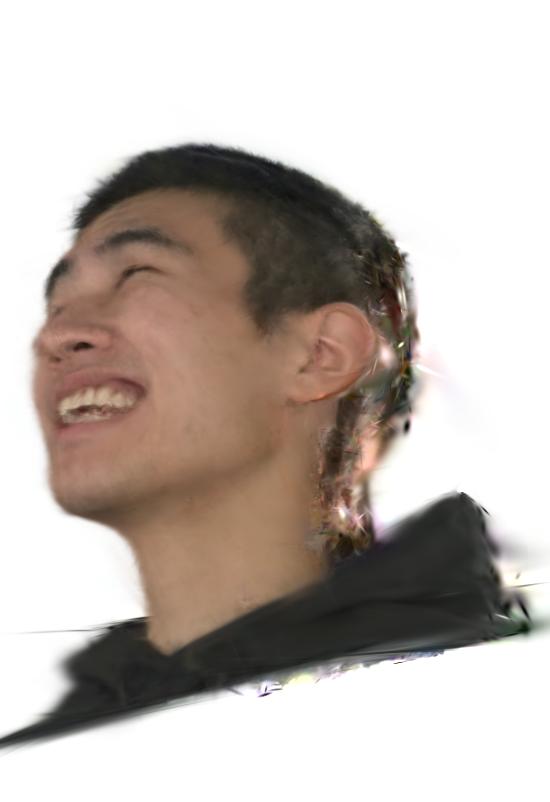} \\
\includegraphics[width=.31\columnwidth]{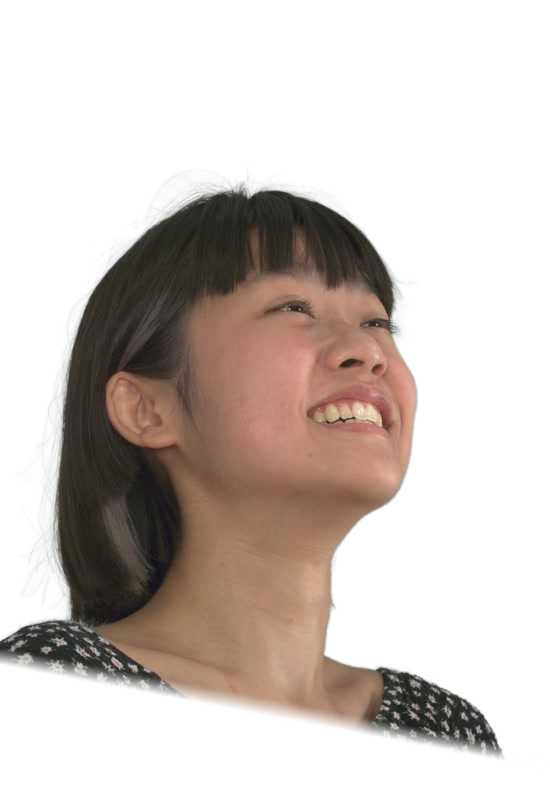} &
\includegraphics[width=.31\columnwidth]{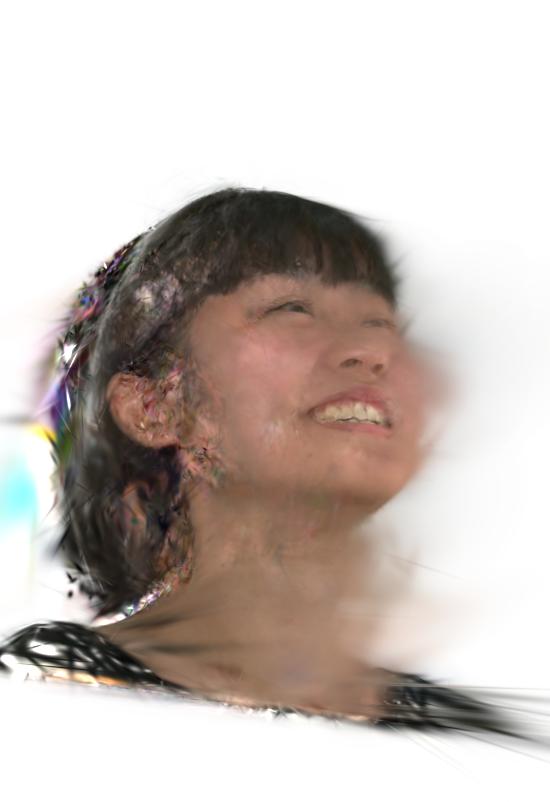} &
\includegraphics[width=.31\columnwidth]{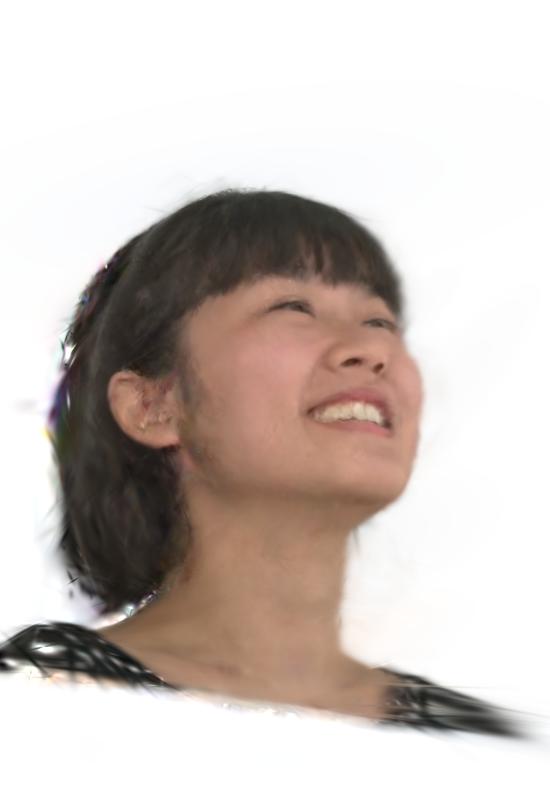} \\
\includegraphics[width=.31\columnwidth]{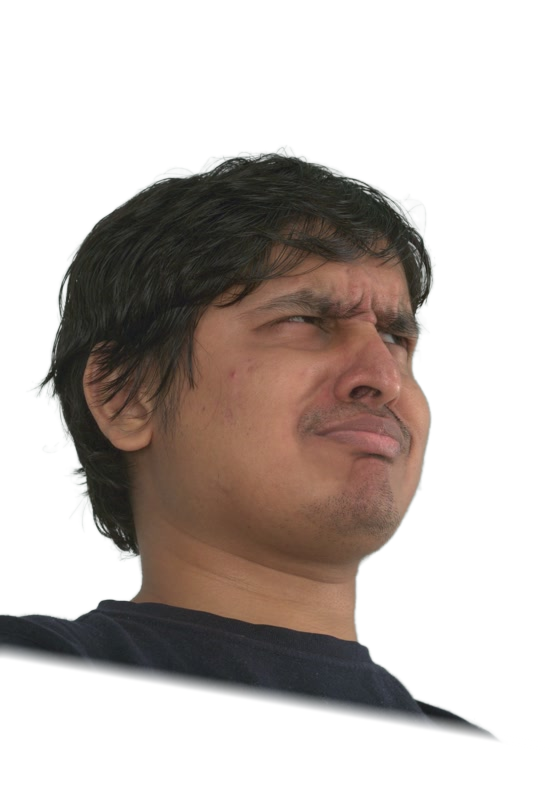} &
\includegraphics[width=.31\columnwidth]{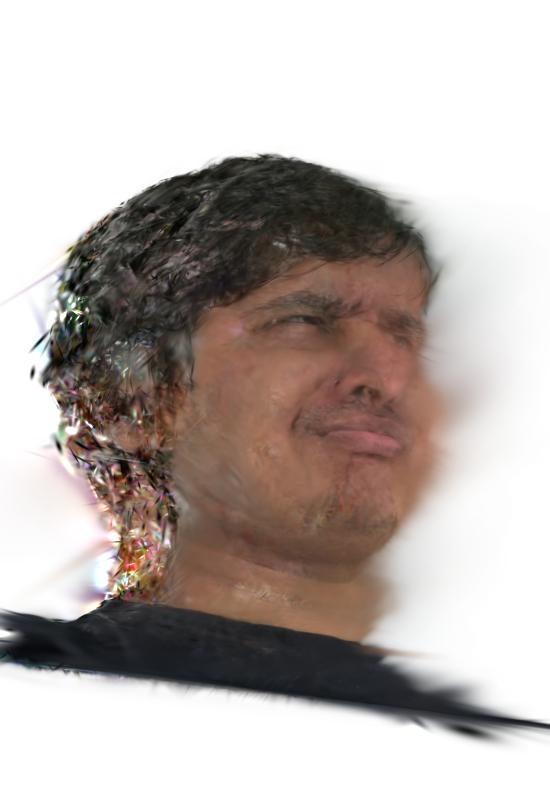} &
\includegraphics[width=.31\columnwidth]{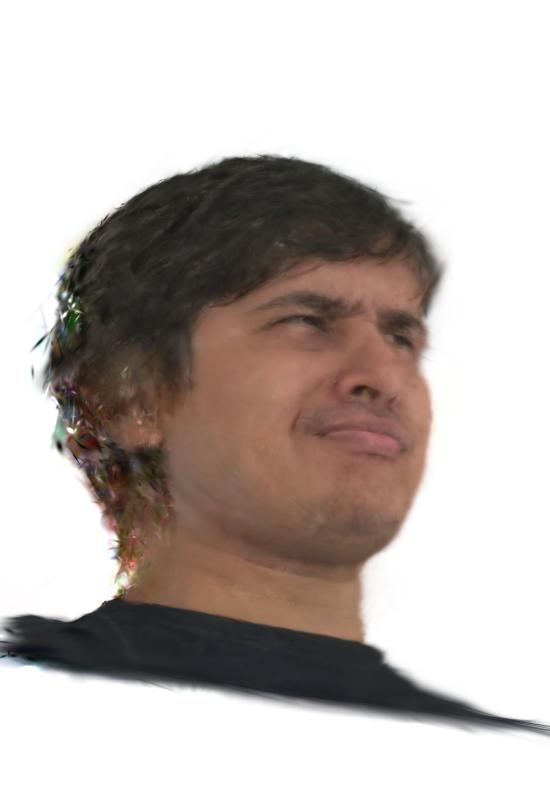}
\end{tabular}
\caption{\newtext{Large-yaw with optional open-mouth examples for RGBAvatar with four input videos. Rows show examples with absolute yaw angles of $37.4^\circ$, $37.7^\circ$, and $36.3^\circ$, respectively. Examples are selected near each identity's median ground-truth yaw and mouth opening within the combined subset.}}
\label{fig:pose-expression}
\end{figure}

\newtext{EDGAR reduced both foreground errors for RGBAvatar and GaussianAvatars, including the combined large-pose/open-mouth subset. CAP4D yielded mixed results. These results show that refinement benefits extend to the tested large-pose and open-mouth conditions for the two video-based pipelines. Figure~\ref{fig:pose-expression} shows examples from three RGBAvatar identities.}

\begin{figure}[t]
\vspace{-2mm}
  \includegraphics[width=\linewidth]{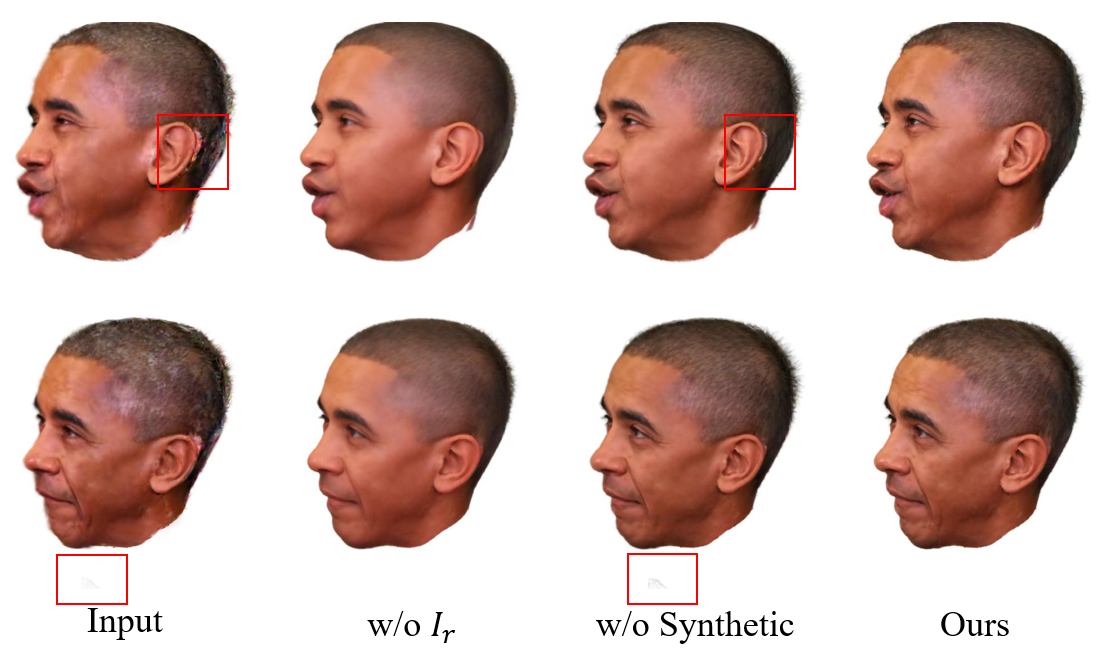}\vspace{-2mm}
  \caption{Ablation study results for $I_r$ and Synthetic data. Some of the artifacts in the hair, ears, and floating blobs exhibited by the input Gaussian avatar are mitigated by the ablated variants. Nonetheless, the variant without $I_r$ produces altered details and overly smooth textures due to the absence of reference information, while the variant without synthetic data retains artifacts such as distortions around the ears and floating Gaussian blobs. In contrast, the full model better preserves identity while effectively removing artifacts and improving overall visual quality. Visible artifacts in the ear and floating Gaussian blobs are highlighted in red boxes.}
  \label{fig:ablation_base}
  \vspace{-2mm}
\end{figure}

\subsection{Ablation Study}\label{sec:ablation}
To better understand the contribution of each component in EDGAR, we conducted an ablation study using CAP4D as the base reconstruction method. Specifically, we evaluated: (1) \textit{w/o $I_r$}, which removes the reference image during refinement; (2) \textit{w/o Synthetic}, which excludes the synthetic split of EDGAR-SET during training; (3) \textit{w/o $\mathcal{L}_{exp}$}, which removes the expression loss; (4) \textit{w/o MoE}, which removes the MoE architecture from the EDGAR denoiser; and (5) \textit{w/ two Experts}, which uses only two experts to examine whether the benefit of MoE simply results from separating the real and synthetic data in EDGAR-SET.

\begin{figure}[t]
  \includegraphics[width=\linewidth]{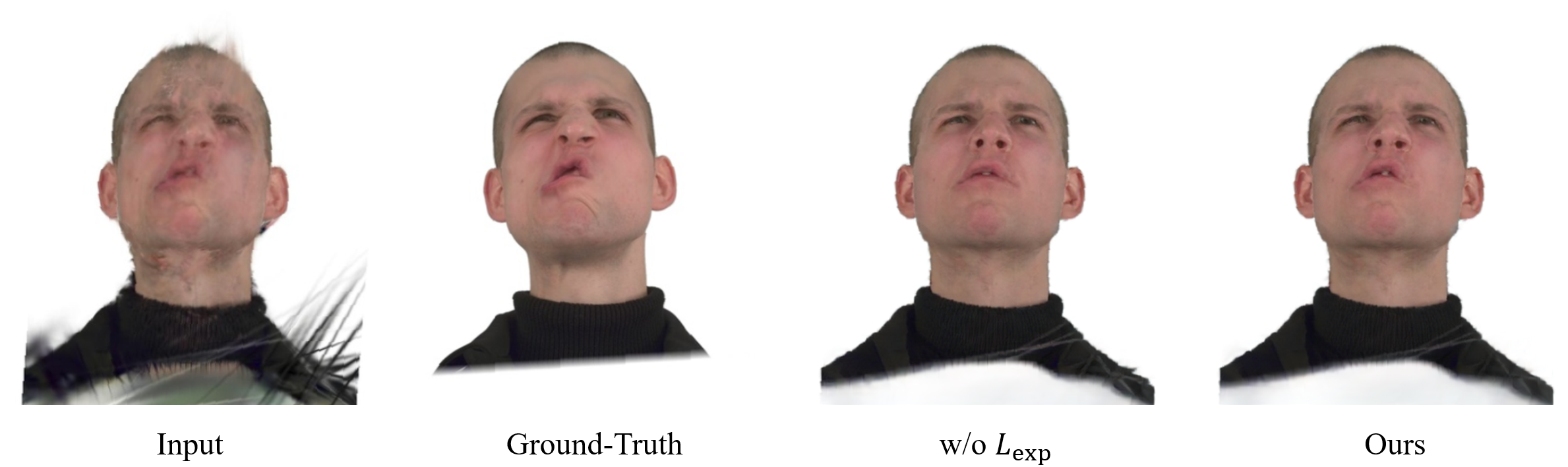}
\caption{Ablation study results on $\mathcal{L}_{exp}$. Removing $\mathcal{L}_{exp}$ degrades the preservation of facial expression details.}  \label{fig:ablation_exp}
\end{figure}
\begin{figure}[t]
  \includegraphics[width=\linewidth]{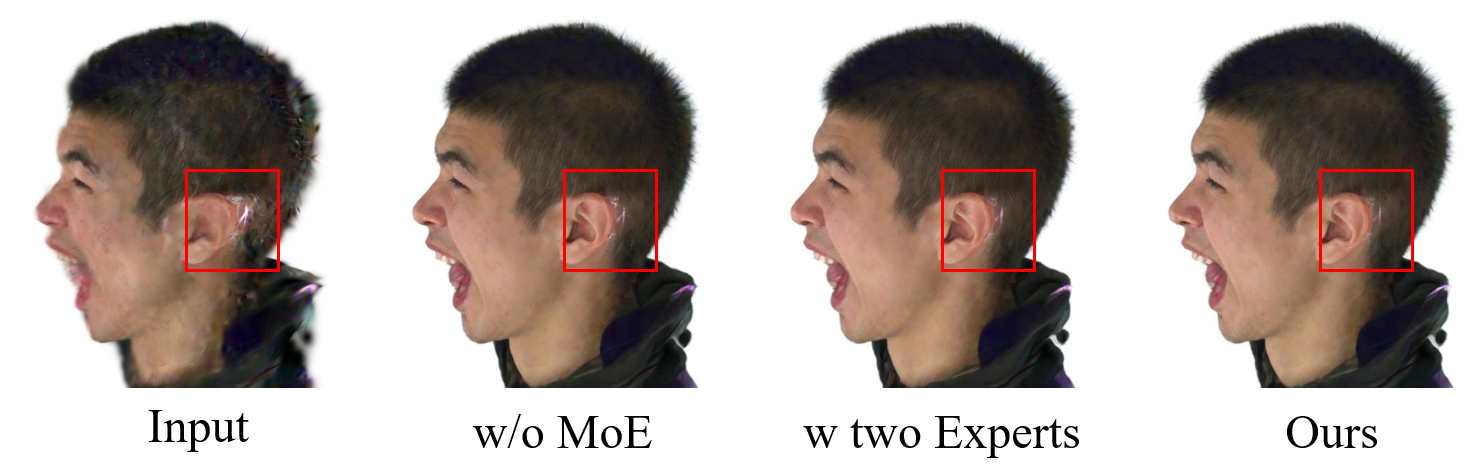}
  \caption{\revisedtext{Qualitative results for the MoE ablation. All variants reduce artifacts, but the variants without MoE and with two experts retain sharp white artifacts around the ears. The full model suppresses these artifacts more effectively in this example.}}
  \label{fig:ablation_moe}
\end{figure}

As summarized in Tables~\ref{tab:ablation} and~\ref{tab:cross-ablation}, corresponding to results from \revisedtext{NeRSemble} and INSTA, respectively, the full EDGAR model achieved the best overall balance between reconstruction fidelity and identity preservation. The ablation results highlight the importance of the reference conditioning term $I_r$ and the synthetic split of EDGAR-SET. Removing $I_r$ led to the largest performance drop in both self-reenactment and cross-reenactment, showing that the reference image and its corresponding Sapiens feature are crucial for preserving identity and stabilizing refinement. Similarly, excluding the synthetic split consistently degrades performance, especially in terms of CLIP, LPIPS, and cross-reenactment CSIM, demonstrating that the synthetic split provides important complementary supervision for generalization. This is also evident in the qualitative comparisons shown in Figure~\ref{fig:ablation_base}, where the full model reduced artifacts and preserved identity more reliably than the variants without $I_r$ or synthetic training data.

The effect of the expression loss $\mathcal{L}_{exp}$ was also meaningful. In self-reenactment, the full model consistently achieved the best CLIP similarity and LPIPS scores across all input view settings. Although the variant without $\mathcal{L}_{exp}$ slightly improved L1 and PSNR, its noticeably worse CLIP and LPIPS scores indicate that optimizing only for pixel-level fidelity does not necessarily lead to better perceptual quality or identity consistency. This tendency becomes more evident in cross-reenactment, where removing $\mathcal{L}_{exp}$ caused a substantial drop in CSIM scores across all input-view settings. Figure~\ref{fig:ablation_exp} shows that the full model better preserves expression details compared to the variant without $\mathcal{L}_{exp}$.

\revisedtext{Finally, the comparison among \textit{w/o MoE}, \textit{w/ two Experts}, and the full model shows that the six-expert model provides improvements in CLIP and in 4-view and 8-view cross-reenactment CSIM, while matching or slightly trailing the ablated variants on other entries. Figure~\ref{fig:ablation_moe} illustrates one case in which the full model better suppresses artifacts around the ear. The reference image, synthetic training split, and expression loss have larger effects than the MoE configuration.}

\subsection{User Study}\label{sec:user-study}
To further assess the perceptual quality of refined avatars, we conducted a web-based user study with 15 participants, with an average age of 27.6 years, including 7 males and 8 females. Each participant was presented with 20 randomized pairs of avatar renderings shown as video sequences generated by the baseline methods (CAP4D and CAP4D+SDEdit) and our method, EDGAR. Each pair consisted of cross-reenactment results, along with the source identity as a reference for identity comparison, and were displayed side-by-side in randomized order to avoid positional bias. Participants were asked to choose their preferred result with respect to three criteria: animation quality, identity preservation, and overall avatar preference.

As shown in Table~\ref{tab:user}, EDGAR was consistently preferred over both baselines across all evaluation criteria. Compared with CAP4D, EDGAR achieved preference rates of 87.33\% in animation quality, 91.33\% in identity preservation, and 96.67\% in overall avatar preference. The preference margin was even larger when compared with CAP4D+SDEdit, where EDGAR achieved 94.67\%, 98.0\%, and 98.0\% for the three criteria, respectively. These outcomes align with our quantitative findings, indicating that EDGAR not only improves animation quality and identity preservation, but also yields avatars that are perceived by participants as more visually convincing overall.

\begin{table}[t]
\centering
\caption{Selection ratio of our method over the baselines in the user study.}
\resizebox{\columnwidth}{!}{%
\setlength{\tabcolsep}{4pt}
\begin{tabular}{lccc}
\toprule
Method & Animation & Preservation & Preference \\
\midrule
Ours vs CAP4D & 87.33\% & 91.33\% & 96.67\% \\
Ours vs CAP4D+SDEdit & 94.67\% & 98.0\% & 98.0\% \\
\bottomrule
\end{tabular}
}
\label{tab:user}
\end{table}



\section{Conclusion}\label{sec:conclusion}
\revisedtext{We introduced EDGAR, a diffusion-driven refinement framework for Gaussian head avatars. EDGAR adapts single-step image refinement to animatable heads through paired avatar-failure data, subject-reference conditioning, expression-aware supervision, and an auxiliary MoE module. Experiments on CAP4D, RGBAvatar, and GaussianAvatars demonstrate improvements in perceptual and reconstruction quality, with gains depending on the starting avatar and evaluation metric.}
\newtext{The shared-avatar update retains approximately the Base rendering rate and improves calibrated cross-view consistency for CAP4D and GaussianAvatars. The pose-and-expression evaluation shows benefits under large yaw and open-mouth conditions.}

\paragraph{Limitations and Future Work.}\label{sec:limitations}
\revisedtext{EDGAR is trained on structurally recoverable degradations: EDGAR-SET filters source--target pairs with large LPIPS differences. Consequently, severe structural errors and artifacts in unobserved regions can remain after refinement, as illustrated in Figure~\ref{fig:comp_self}. Recovering such failures may require additional observations or geometric supervision.}

\newtext{Independently refined pseudo-targets can disagree across views, causing the shared Gaussian representation to average conflicting details or retain view-dependent errors. Although our cross-view evaluation shows that EDGAR generally improves 3D consistency, explicit multi-view constraints and confidence-aware selection of refined targets could further reduce such inconsistencies. Mouth interiors and translucent appearance also remain challenging. In addition, the temporal evaluation shows that the mean tWE decreased for CAP4D and RGBAvatar but increased for GaussianAvatars, indicating that improvements in spatial fidelity do not necessarily translate into greater temporal stability across all reconstruction pipelines}. Explicit temporal supervision and improved modeling of dynamic facial regions are promising directions for future work.


\begin{acks}
This work was supported by Institute of Information \& Communications Technology Planning \& Evaluation(IITP) grant funded by the Korea government(MSIT) (No.RS-2024-00439499, Generating Hyper-Realistic to Extremely-stylized Face Avatar with Varied Speech Speed and Context-based Emotional Expression)
\end{acks}

\bibliographystyle{ACM-Reference-Format}
\bibliography{main}


\end{document}